\documentstyle[12pt,epsfig]{article}
\begin{document}

\begin{titlepage}
\rightline{April 2013}
\vskip 2cm
\centerline{\Large \bf
Galactic structure explained 
}
\vskip 0.4cm
\centerline{\Large \bf
with dissipative mirror dark matter
}

\vskip 2.2cm
\centerline{R. Foot\footnote{
E-mail address: rfoot@unimelb.edu.au}}

\vskip 0.7cm
\centerline{\it ARC Centre of Excellence for Particle Physics at the Terascale,}
\centerline{\it School of Physics, University of Melbourne,}
\centerline{\it Victoria 3010 Australia}
\vskip 2cm
\noindent
Dissipative dark matter, such as mirror dark matter and related 
hidden sector dark matter candidates,
require an energy source to stabilize dark matter halos in spiral galaxies. 
It has been proposed previously that supernovae could be the source of this energy.
Recently, it has been argued that this mechanism might explain two
galactic scaling relations inferred from observations of spiral galaxies. One of which is that 
$\rho_0 r_0$ is roughly constant, and another relates the galactic luminosity to $r_0$.
[$\rho_0$ is the dark matter central density and $r_0$ is the core radius.]
Here we derive equations for the heating of the halo via supernova energy, and the 
cooling of the halo
via thermal bremsstrahlung. These equations are numerically solved to obtain
constraints on the $\rho_0, \ r_0$ parameters appropriate for spiral galaxies. 
These constraints are in remarkable agreement with the aforementioned scaling
relations.

\end{titlepage}


\section{Introduction}

The standard $\Lambda$CDM scenario\cite{rev1}, 
which invokes weakly interacting dark matter particles, 
has proven to be extremely successful in explaining the observed large scale structure 
and cosmic microwave background anisotropies\cite{cmbpapers}.
However, this scenario is challenged by observations on smaller scales.
For example,
measurements of rotation curves of spiral galaxies indicate the
existence of a dark
matter cored profile (e.g.\cite{core}) in disagreement with the cuspy profile
predicted by simulations of non-interacting dark matter\cite{nfw}.

Another challenge to the standard $\Lambda$CDM scenario is the lack
of any evidence for new stable particles in LHC collider searches\cite{lhc}.
From a particle physics perspective, though, there is no compelling reason to
favour weakly interacting particles over richer dark matter scenarios.
In fact,
the success of the standard model in explaining all
collider data to date is a definite hint that dark matter resides in a
hidden sector.  That is, the fundamental Lagrangian decomposes into two
sectors, one describing the standard particles and forces, and another
which will contain the dark matter:
\begin{eqnarray}
{\cal L} = {\cal L}_{SM} + {\cal L}_{dark}
\ .
\end{eqnarray}
The sector describing the ordinary particles has $G = SU(3) \otimes
SU(2)_L \otimes U(1)_Y$ gauge symmetry, while the dark sector has
independent gauge symmetries, $G'$.
In this case one expects dark matter to be multi-component, 
self-interacting and perhaps dissipative, if the hidden sector has an
unbroken $U(1)'$ gauge symmetry. Such scenarios can also successfully
explain the large scale structure and CMB, but can yield very different
physics on small scales. Such hidden sector dark matter has been discussed in the literature
in a variety of contexts, see. e.g.\cite{blin, hodges, review, lss, fv, hall, h1, h2, mohap,china,china2,hamed,feng,ring,ham}.

If the hidden sector does indeed contain an unbroken $U(1)'$ gauge symmetry, then
the associated `dark' photon can kinetically mix with the ordinary photon:
\begin{eqnarray}
{\cal L}_{mix} = \frac{\epsilon}{2} F^{\mu \nu} F'_{\mu \nu}
\label{kine}
\end{eqnarray}
where $F_{\mu \nu}$ ($F'_{\mu \nu}$) is the field strength tensor for the photon (dark photon).
Such kinetic mixing is gauge invariant and renormalizable\cite{he} with $\epsilon$ 
viewed as a fundamental
parameter of the theory.
The physical effect of the kinetic mixing interaction\cite{holdom} is to induce a tiny ordinary
electric charge ($\propto \epsilon$) for the hidden sector $U(1)'$ charged particles.

Mirror dark matter is a well motivated and concrete example of hidden
sector dark matter
in which ${\cal L}_{dark}$ is an exact duplicate of the standard model
sector\cite{flv}. 
This means that the
hidden sector has gauge symmetry $G' = SU(3)'\otimes SU(2)'_L \otimes U(1)'_Y$. 
If the chiral left and right handed fermion fields are interchanged in such a hidden sector
then the theory has an unbroken `mirror' symmetry mapping each ordinary
particle onto a `mirror' particle (along with $x \to -x$). 
That is, for each ordinary particle, $e, \nu, u, d, ..., \gamma,...$ there is 
a corresponding mirror particle, which we denote with a prime ($'$): $e', \nu', u', d', ..., \gamma',...$.
The unbroken mirror symmetry 
ensures that the mirror particles have the same masses as their corresponding ordinary counterparts.
Similarly the gauge self-interactions (mirror electromagnetism etc) have the same form
and strength in the mirror sector as they do in the ordinary sector.
Although the
focus here is on mirror dark matter, our results may be relevant to closely
related hidden sector models, such as the ones discussed in ref.\cite{china}.

Mirror dark matter has emerged as an interesting dark matter candidate, 
for reviews and more complete bibliography see e.g. \cite{review}. 
Mirror dark matter can explain\cite{lss} the
large scale structure of the Universe - the matter power spectrum and CMB - in a manner analogous 
to standard collisionless cold dark matter models provided that $\epsilon \stackrel{<}{\sim} 10^{-9}$\cite{cmb}.
Such values of $\epsilon$ are well consistent with direct laboratory limits, which arise 
from rare decays of orthopositronium\cite{glashow}.  Importantly,
mirror dark matter can also explain\cite{foot} the positive dark matter signals from the 
DAMA\cite{dama} experiment along with the more tentative signals from 
CoGeNT\cite{cogent}, CRESST-II\cite{cresst} 
and CDMS/Si\cite{cdms}
direct detection experiments. This explanation requires photon-mirror photon kinetic mixing
of strength $\epsilon \sim 10^{-9}$. 
Of course, this type of rich dark matter candidate can feature a 
whole range of phenomena,  
especially on small scales (galactic and smaller) which have only begun to be explored 
(see e.g.\cite{novel,macho1,sph,sil,p88,fsnew}). 

In this article we return to the problem of galaxy structure in the context 
of mirror dark matter.
The existence of the unbroken $U(1)'$ interaction will inevitably lead to
significant self-interactions of the mirror particles. 
An implication of this is that
galactic halos of spiral galaxies would have to be composed (predominately) of mirror 
particles in a pressure supported spherical plasma\cite{sph}. 
There may also be a subcomponent 
consisting of compact objects such as old mirror stars.
Some of the interactions within the halo will be dissipative, 
such as thermal bremsstrahlung (e.g. $e' + He' \to e' + He' + \gamma'$)
which can cool the halo.
At first sight, this might put into question the very existence of the halo.
At any rate, the stability of the halo needs to be explained.
A possible explanation was
suggested sometime ago\cite{sph}. The idea is that ordinary
core collapse supernovae provide the required energy.
In the hot and dense core of a type II supernova mirror electrons and 
positrons can be created from
kinetic mixing induced 
plasmon decay processes\cite{raffelt}.
Thus ordinary supernovae can be a source of light mirror particles
as well as the ordinary neutrinos.
Indeed, it is estimated that mirror particles 
($e', \ \bar e', \ \gamma'$) 
carry off
roughly half of the core collapse supernova energy
if $\epsilon \sim 10^{-9}$
\cite{raffelt,sil}. 
A significant fraction 
of this energy can be absorbed by the mirror particle halo and thereby
potentially
replace the energy lost due to dissipative interactions. 
Order of magnitude  estimates\cite{sph} suggested that the amount
of heat generated roughly matched the energy
dissipated for the Milk Way. The same mechanism could be responsible for
stabilising the halos in all spiral galaxies.  If so, then matching the total
heat supplied from ordinary supernovae to the energy dissipated 
suggests\cite{r69}
a rough galactic scaling relation for spirals: $R_{SN} \propto \rho_0^2 r_0^3$. Here,
$\rho_0, r_0$ are the dark matter central mass density and core radius and
$R_{SN}$ is the galactic supernova rate. It
was further suggested that
the way in which energy from supernovae is distributed 
would lead to a cored dark matter distribution. A
second scaling relation, $\rho_0 r_0 \approx constant$ might
be explained in this way, it was argued\cite{r69}.

The purpose of this article is to provide a more detailed numerical
analysis of this whole picture.
Let us mention at the outset that the problem of small scale structure
is a complicated one and the analysis performed here, although 
progress over the rough physical arguments of ref.\cite{sph,r69}, still has some deficiencies. 
A number of assumptions are made, which would require
further checks and refinements. More importantly, we do not attempt to evolve the galaxy
from an early time to its present state, but rather see if we can
at least explain some of the current properties of spiral galaxies. 

This paper is structured as follows. In section 2 we give a brief 
overview of the central idea, that dissipative dark matter candidates like
mirror dark matter (and by extension, closely related hidden sector models),
can have their halo's stabilized via supernova energy.
In section 3 we discuss the hydrostatic equilibrium condition, which we numerically
solve to obtain the galactic temperature profile for several example spiral
galaxies.
In section 4 we consider the ionization state of the halo. The equations
governing the ionization fractions of the mirror helium, mirror hydrogen
and mirror metal components are given and numerically solved.
In section 5 we 
derive equations for the heating of the halo via supernova energy, and the cooling of the halo
via thermal bremsstrahlung. These equations are numerically solved to obtain
constraints on the $\rho_0, \ r_0$ parameters appropriate for spiral galaxies. 
These constraints are then compared with the scaling
relations inferred from observations of galactic rotation curves.
In section 6, we comment on dwarf spheroidal galaxies and
in section 7 we discuss briefly elliptical galaxies and galaxy clusters.
Finally we give a few concluding remarks in section 8.



\section{The heating of the galactic halo}  

The physical picture is that spiral galaxies such as the Milky Way are currently 
composed of ordinary
matter in a disk, and mirror dark matter predominately in a (roughly) spherical halo.
The halo consists of an ionized plasma 
formed out of the 
mirror particles, $e', H', He', O', Fe', ...$. 
The plasma dissipates energy due to thermal bremsstrahlung and other processes and this
energy needs to be replaced.  The idea\cite{sph} is that ordinary supernovae can supply this
required energy if photon-mirror photon kinetic mixing exists, Eq.(\ref{kine}).
Such kinetic mixing gives the mirror electron and positron a tiny ordinary electric charge 
of magnitude $\epsilon e$. 
The energy loss rate for production of such minicharged particles from supernovae has been
estimated in Ref.\cite{raffelt}:
\begin{eqnarray}
Q_P = {8 \zeta_3 \over 9 \pi^3}  \epsilon^2 \alpha^2 \left(
\mu^2_e + {\pi^2 T^2
\over 3}\right) T^3 Q_1
\label{raf1x}
\end{eqnarray}
where $Q_1$ is a factor of order unity, and $\mu_e$ is the electron chemical potential and 
$T \approx 30$ 
MeV is the temperature of the supernova core.
Demanding that 
$Q_P$ does not exceed the energy loss rate due to neutrino
emission implies that 
$\epsilon \stackrel{<}{\sim} 10^{-9}$\cite{raffelt}. 
Thus,
supernova can be a source of energetic light mirror particles 
$e', \bar e', \gamma'$ 
which can ultimately replace the energy lost due to radiative cooling.
This heating of the halo is in the central region of the galaxy, which leads to
the temperature having a mild radial dependence. The halo is generally hotter at the center and
decreases as the distance from the center, $r$, increases.
The `average' halo temperature (say at a distance $r=3r_D$ where $r_D$ is the disk scale length)
is typically of order 300 eV for the Milk Way, and
ranges from $10$ eV for the smallest spirals to around few keV
for the largest spirals (see the discussion in the following section).

The amount of supernova energy required to replace the halo energy lost due to radiative cooling
is sizable. Estimates\cite{sph,r69} indicate that at least  
a few percent of the total supernova energy needs to be absorbed by the halo. 
Is this reasonable?
Let us assume a kinetic mixing parameter $\epsilon \sim 10^{-9}$, so that around half of type II supernova
energy is converted into $e', \ \bar e', \gamma' $ emitted from the core initially with energies $\sim $ MeV.
The huge number of energetic $e', \bar e', \gamma'$ particles injected into the region [($\sim 1\ {\rm pc})^3$]
around ordinary supernova will rapidly cool, ultimately converting most of their energy into mirror 
photons.
The energy spectrum of these mirror photons is naturally very difficult to predict 
but it could have some vague resemblance to the 
$\gamma$ spectrum of ordinary Gamma Ray Bursts (GRB's). Recall GRB's
feature a fairly wide spectrum of energies with mean $\sim$ 700 keV with a few percent of energy radiated below 
around 10 keV.
In any case,
these mirror photons will then heat the mirror particle halo, potentially supplying the energy lost from
the halo due to radiative cooling. 

It has been argued previously\cite{sph,r69} that this $\gamma'$ energy cannot be transferred to the halo via elastic (Thomson)
scattering off free $e'$ in the plasma.
The Thomson cross-section is at least an order of magnitude too small. Thus,
if the halo contains only $H'$ and $He'$ components, then it is hard to see how
enough energy can be absorbed by the halo to replace the energy lost due to radiative cooling.
However if the halo
contains mirror metal components then the situation is much more promising.
The heavy metal components are not fully ionized but can have their atomic K-shells filled.
The photoionization cross-section is many orders of magnitude larger than the Thomson cross-section and
even a small metal component can make the halo optically thick, at least for a range of $\gamma'$ energies.
Once the energetic K-shell $e'$ is ejected from the ion, it will interact with the free $e'$ and the ions
in the vicinity (typically $\sim $ pc) and thermalize. 

The total photoelectric cross-section\footnote{Unless otherwise specified, 
we use natural units with $\hbar = c = 1$.}
of a mirror element with atomic number, $Z$, is given by (see e.g.\cite{book5}):
\begin{eqnarray}
\sigma_{PE} (E_{\gamma'}) = 
{g 16\sqrt{2} \pi \over 3m_e^2 } \alpha^6  Z^5 \left[ {m_e \over E_{\gamma'}
}\right]^{7/2} 
\ \ {\rm for} \ E_{\gamma'} \gg I
\label{pe4}
\end{eqnarray}
where $I$ is the $e'$ binding energy and $g = 1$ or 2 counts the number 
of K-shell mirror electrons present.
Evidently, the photoelectric cross-section
decreases with mirror photon energy like $(E_{\gamma'})^{-7/2}$.
For $E_{\gamma'}$ near threshold the cross-section has a 
slightly softer behaviour, $\sigma_{PE} \propto 1/E_{\gamma'}^3$ and drops abruptly to zero at 
$E_{\gamma'} = I$\cite{bookastro}.
The contribution to the optical depth 
from such inelastic scattering for $\gamma'$ propagating out from the galactic center is 
\begin{eqnarray}
\tau_{IS} &= & \sum_{A'} 2\int_0^{\infty} \sigma_{PE} n_{A'} dr 
\nonumber \\
&\sim & \sum_{A'} 2\rho_0 r_0 \sigma_{PE} \left[ {\xi_{A'} \over m_{A'}}
\right]
\end{eqnarray}
where $\xi_{A'}$ is the proportion by mass of the mirror 
metal component, $A'$ (e.g. $A' = C', O', Si', Fe',...$).
The quantities $\rho_0$ and $r_0$ are the halo central mass density and core radius, whose
product $\rho_0 r_0$ has been inferred to be roughly constant (i.e. independent of galaxy
luminosity), with value around $10^{2.2} m_{\odot}/{\rm pc}^2$ for a Burkert
profile\cite{kf,sal2,donato}.

If we consider just the $Fe'$ component,
we find that the optical depth is substantial, $\tau_{IS} \stackrel{>}{\sim} 0.1$, provided  
\begin{eqnarray}
E_{\gamma'} \stackrel{<}{\sim} 30 \ {\rm keV} \ \left[
{\rho_0 r_0 \over 10^{2.2} m_\odot/{\rm pc}^2}\right]^{2/7} \left[{\xi_{Fe'} \over 0.05} \right]^{2/7}
\ .
\end{eqnarray}
If the $\xi_{Fe'}$ component is not too small ($\stackrel{>}{\sim} 0.01$) and
assuming that 
the supernova $\gamma'$ spectrum peaks above $\sim 30$ keV and falls sharply at low energy (like a thermal
spectrum) then the supernova energy being absorbed by the halo will arise (predominately) from
$\gamma'$ in the energy range
\begin{eqnarray}
9 \ {\rm keV} \stackrel{<}{\sim} E_{\gamma'} \stackrel{<}{\sim} 30\ {\rm keV}\ . 
\end{eqnarray}
Even if there are more abundant lighter components, such as $O'$, this may  not change
this picture greatly. Thus, including 
just the $Fe'$ component, might be sufficient, when considering how much supernova energy
is being absorbed by the halo. We will see in section 5 that the derived galactic scaling
properties of spiral galaxies are relatively insensitive to the precise details of the
supernova $\gamma'$ spectrum.

To summarize, ordinary core collapse supernova will produce light mirror particles, $e', \bar e', \gamma'$ from their
core with total energy comparable to the neutrino burst provided that kinetic mixing of strength $\epsilon \sim 10^{-9}$
exists.
The bulk of this energy is expected to be converted into mirror photons, $\gamma'$, in the region around supernova.
The details of the resulting $\gamma'$ energy spectrum are poorly understood, 
but only the part of this spectrum below 
around 30 keV will be important for heating the halo. 
This heating is achieved by interactions (photoionization) with heavy mirror
metal components, which occurs because these components 
retain their K-shell mirror electrons.
Considering just $Fe'$ might be sufficient, as far as the heating of the halo is concerned,
provided that the proportion of the supernova $\gamma'$ energy 
spectrum below the $Fe'$ K-shell binding energy, $\approx$ 9 keV
is small.


\section{Hydrostatic equilibrium}

The halo has two components, a plasma component and a dark disk/compact object component.
Microlensing observations\cite{macho} provide some evidence that
the mass of the plasma component dominates
over that of the compact object component, and we henceforth focus on the plasma component.

The plasma component consists of a set of  mirror particles, $e', H', He',...$.
What is the {\it current} chemical composition of the halo?
Early Universe cosmology suggests\cite{paolo1} that 
the primordial mirror helium mass fraction is around $Y^P \approx 0.9$ for $\epsilon \sim 10^{-9}$,
with negligible primordial production of mirror metal components.
In the first billion years or so, substantial mirror star formation and evolution
is possible (currently, though, the halo is far too hot for much mirror star formation
to occur). During this early period mirror metals could have been produced reasonably efficiently
given that  mirror stars with large mirror helium mass fraction
evolve 10-100 times faster than ordinary stars (which have $Y^P \approx 0.25$)\cite{bersar}.
Presumably this early epoch of mirror star formation is responsible for the 
current
halo metal component (likely at least $\xi_{Fe'} \stackrel{>}{\sim} 0.01$)
required for the halo to absorb 
enough of the mirror photons produced in the hot region around ordinary supernovae\footnote{
The rampant mirror star formation and evolution in the first few billion years or so of
galactic evolution presumably included a large number of mirror supernovae.
If the kinetic mixing interaction exists, then mirror supernovae should be a source of
a large ordinary x-ray photon flux for the same reasons that 
ordinary supernovae are suspected to be a source
of a large mirror x-ray photon flux. One could even speculate that 
these photons may have been responsible for the
reionization of ordinary matter inferred from the CMB observations\cite{cmbpapers}.
Importantly, the huge photon flux may not prevent the collapse of ordinary matter
onto a disk because the ordinary matter with negligible metal component (at that time)
would absorb a relatively small fraction of the mirror supernovae energy.
}.
A significant halo metal component is also inferred from the mirror dark matter explanation\cite{foot} of the 
DAMA\cite{dama}, CoGeNT\cite{cogent} and CRESST-II\cite{cresst} 
direct detection experiments. 
The end result is a halo currently composed primarily of $He', H'$, with $He'$ mass
fraction around $0.9$. Additionally there is
a small metal fraction (a few percent by mass), which we take to be $Fe'$. 
An important quantity is the
mean mass of the plasma particles:
\begin{eqnarray}
\bar m \equiv \sum m_{A'} n_{A'}/\sum n_{A'} 
\label{mbar}
\end{eqnarray}
where $A' = e', H', He',Fe'$ and  $n_{A'}$ is the $A'$ particle number density.
For a fully ionized plasma, we estimate that $\bar m = 1.1$ GeV.

The set of  mirror particles, $e', H', He',...$
in the plasma 
interact with each other via Coulomb scattering. 
These self-interactions suggest that
mirror dark matter forms a pressure supported halo.
At the present time such a halo would be expected to be
in hydrostatic equilibrium, where the force of gravity is balanced by
the pressure gradient. That is,
\begin{eqnarray}
{dP \over dr} &=& -\rho (r) g(r) \nonumber \\
&=& -\bar m n_T (r) {v_{rot}^2 \over r}
\label{p9}
\end{eqnarray}
where $\rho (r) = \bar m n_T (r)$, with $\bar m$
being the mean mass of the matter mirror particles defined in Eq.(\ref{mbar}),
$n_T (r)$ total dark matter particle
number density\footnote{If we were
to consider also a dark disk/compact object component made of old mirror stars, 
mirror white dwarfs etc,
then $n_T (r)$ in Eq.(\ref{p9}) would be just the plasma component of the halo.}
and $g(r)$ is the local acceleration due to gravity. 
Here $P(r) = n_T (r)T(r)$ and $v_{rot} (r)$ is the local rotational velocity.

We shall assume that both the mirror particles and the ordinary baryons
are distributed with spherically symmetry. Obviously in the
central regions of spiral galaxies this might not be a good 
approximation, since the ordinary matter is distributed predominately in a disk.
However so long as we consider $r \stackrel{>}{\sim} r_D$, then the assumption
of spherical symmetry could be reasonable.
Furthermore, since the typical core radius of spirals is inferred to be 
much larger than the disk scale length, $r_D$, it seems reasonable that
the physics responsible for the existence and properties of the dark matter core
will not depend too sensitively on the details of the mass distribution at $r \ll r_0$.

Anyway, with the assumption of spherical symmetry,
$g$ (and hence $v_{rot}^2/r$) can be related to 
the total mass density, $\rho_{total}$, via:
\begin{eqnarray}
g(r) = {G_N\over r^2} \int_0^r  \rho_{total} dV
\end{eqnarray}
where $G_N$ is Newton's constant.
The total mass density, $\rho_{total}$, can be separated into a contribution from ordinary baryons
and that due to mirror particles.
The baryonic contribution of spiral galaxies is approximated by
a Freeman disk with surface density:
\footnote{
We do not include any other baryonic contribution other than the disk, so that $m_D$ represents
the total baryonic mass of the galaxy.}
\begin{eqnarray}
\Sigma = {m_D \over 2\pi r_D^2} \ e^{-r/r_D}
\end{eqnarray}
where $m_D$ is the disk mass 
and $r_D$ is the disk scale length.
Defining a spherically symmetric distribution, $\rho_D$, by requiring 
that the mass within a radius $r$ is the same as that of the 
disk, i.e. $\int^{r}_{0} \rho_D 4\pi r'^2 dr' \equiv \int^r_0 \Sigma \ 2 \pi r' dr' $, 
we have:
\begin{eqnarray}
\rho_D (r) = {m_D \over 4\pi r_D^2 r} e^{-r/r_D}
\label{12z}
\ .
\end{eqnarray}
Studies of spiral galaxies have found that the baryonic mass $m_D$ correlates
with the disk radius, $r_D$ via\cite{lapi,PSS}:
\begin{eqnarray}
\log \left( {r_D \over {\rm kpc}}\right) = 0.633 + 0.379
\log \left( {m_D \over 10^{11} m_{\odot}}\right) 
+ 0.069 \left[\log \left( {m_D \over 10^{11} m_{\odot}}\right)\right]^2
\ .
\label{rd}
\end{eqnarray} 


The philosophy adopted here is that the dark matter distribution within
spiral galaxies is governed by
hydrostatic equilibrium, dissipation and supernova heating. 
At the current epoch, these conditions might be sufficient to determine 
the dark matter density profile, independently of the past history of the galaxy.
In the present work, though, we shall
assume we know something about the form of the dark matter distribution (and justify
this form later by showing that it is an approximate solution to the derived equations).
We 
assume that dark matter can be approximated by a smooth
cored distribution
\begin{eqnarray}
\rho_{dm} &=& \bar m n_T (r) 
\nonumber \\
          &=& {\rho_0 r_0^3  \over (r^2 + r_0^2) (r + r_0)
} 
\label{1}
\end{eqnarray}
where $r_0, \ \rho_0$ are  the dark matter core radius and central density respectively.
Such a distribution, known as the Burkert profile, has been suggested by fits to rotation curves of spiral galaxies,
and other data\cite{bp}. Several scaling relations, have 
been derived for $\rho_0, \ r_0$ from such data fitting. By adopting the same
dark matter profile, we can hope to compare the dynamically derived 
constraints on $\rho_0, \ r_0$ with
the scaling relations found from the data.
In a separate article, we shall examine more general forms of halo dark 
matter distribution suggested by this dynamics\cite{foottoappear}.

If $\rho_{D} (r) $ and $\rho_{dm} (r)$ are known
we can use the hydrostatic equilibrium condition
to figure out the temperature profile, $T(r)$.
To do this, we need a boundary condition. 
Far from the galactic center, i.e. far from heating sources we expect 
isothermal conditions, which motivates
$dT/dr = 0$ at large galactic distance, $R_{gal}$, which we take to be $50 r_D$.
Our numerical results are approximately independent of the particular value
of $R_{gal}$ chosen so long as $R_{gal} \gg r_D$.
To get an idea of the typical temperature profiles that we expect for spiral galaxies,
we consider three examples: (a) a small sized spiral galaxy with disk mass $m_D = 10^9 m_\odot$ 
and dark matter core radius $r_0 = 4$ kpc, 
(b) a medium sized spiral galaxy ($\sim$ Milk Way) with disk mass $m_D = 10^{11} m_\odot$ and
dark matter core radius $r_0 = 12$ kpc and 
(c) a large sized spiral galaxy with disk mass $m_D = 10^{12} m_\odot$ and
dark matter core radius $r_0 = 40$ kpc. 
In each case we take $\rho_0 = 10^{2} (m_{\odot}/{\rm pc}^2)/r_0$, consistent with results inferred 
from observations\cite{kf,sal2,donato}.
The results of numerically solving Eq.(\ref{p9}), with the boundary condition
discussed above are shown in figure 1.
The figure indicates that with a Burkert dark matter density profile the corresponding temperature 
profile derived from the hydrostatic equilibrium condition smoothly rises towards the
central region of the galaxy, where it is roughly isothermal.

\centerline{\epsfig{file=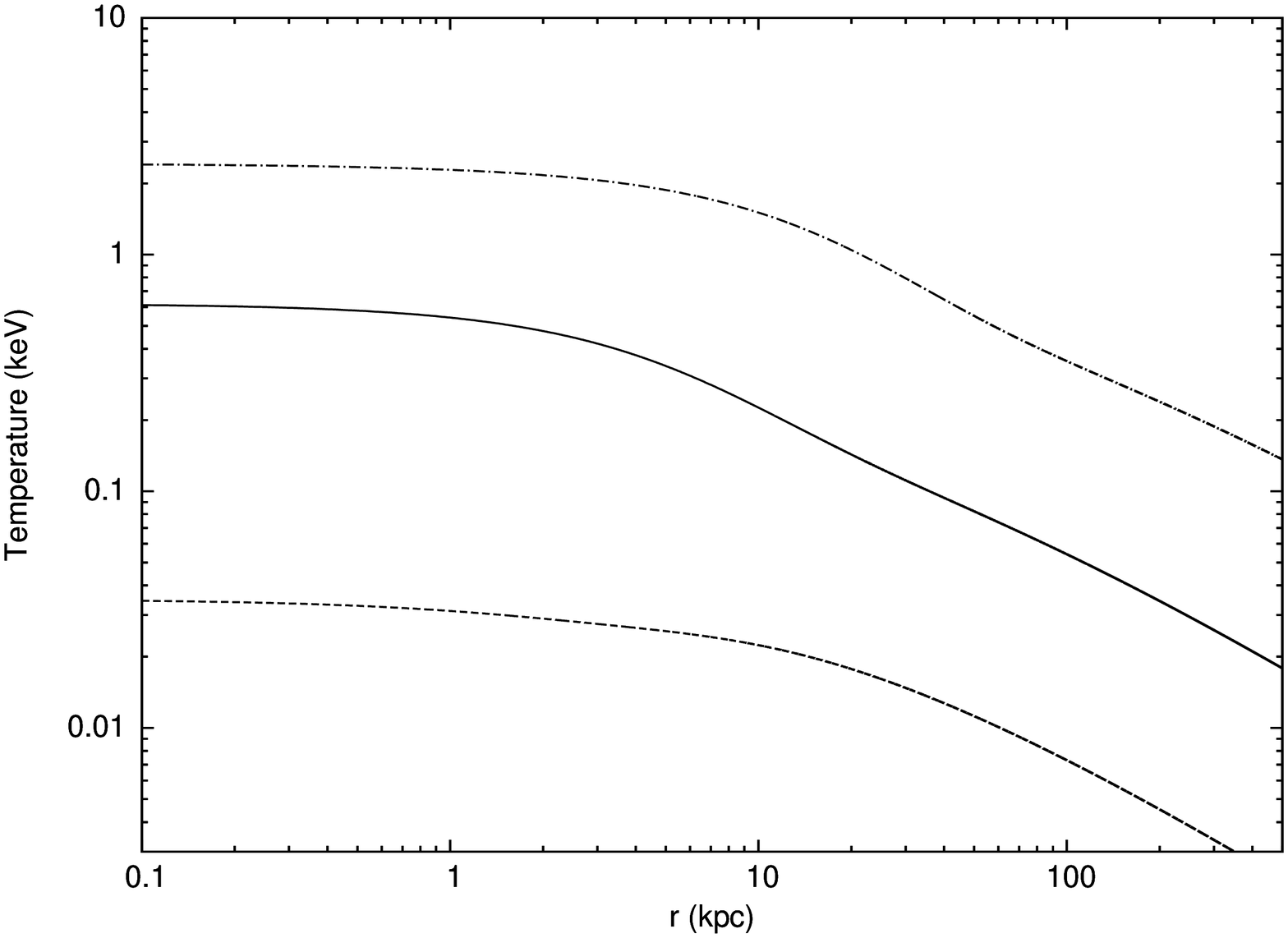, angle=270, width=12.8cm}}
\vskip 0.5cm
\noindent
{\small Figure 1: The temperature profile of the mirror dark matter plasma in 
spiral galaxies for three examples:
(a) a small sized spiral galaxy with disk mass $m_D = 10^9 m_\odot$ 
and dark matter core radius $r_0 = 4$ kpc (dashed line), 
(b) a medium sized spiral galaxy ($\sim$ Milk Way) with disk mass $m_D = 10^{11} m_\odot$ and
dark matter core radius $r_0 = 12$ kpc (solid line) and 
(c) a large sized spiral galaxy with disk mass $m_D = 10^{12} m_\odot$ and
dark matter core radius $r_0 = 40$ kpc (dashed-dotted line). 
In each case we take $\rho_0 = 10^{2} (m_{\odot}/{\rm pc}^2)/r_0$. 
}

\vskip 1.2cm


\section{The ionization state of the halo}

For the temperature range of interest for spiral galaxies, 
typically $0.01 \ {\rm keV} \stackrel{<}{\sim} T \stackrel{<}{\sim}$  few keV,
the plasma is kept ionized by $e'$ collisions.
Considering, just the mirror helium component for now,
the relevant processes are:
\begin{eqnarray}
 e' + {He'}^{0} &\rightarrow & {He'}^{+} + e' + e' 
\nonumber \\
 e' + {He'}^{+} &\rightarrow & {He'}^{2+} + e' + e' 
\label{14x}
\end{eqnarray}
where ${He'}^0, {He'}^+, {He'}^{++}$ denote the neutral mirror helium atom,
singly charged mirror helium ion 
and doubly charged mirror helium ion.
Since the Lagrangian describing the particle physics of the mirror
particles is exactly analogous to the one describing the ordinary
particles and forces, the cross-section for the above processes is
precisely the same as for the corresponding ordinary particle process.
The cross-section for these processes 
is known to be reasonably well approximated by the Lotz formula\cite{lotz}:
\begin{eqnarray}
\sigma_I = 4.5\times 10^{-14} \left[ {ln (E/I) \over EI/eV^2}\right] \ {\rm cm^2}
\label{ion}
\end{eqnarray}
where $E \ge I$ is the energy of the incident $e'$ and $I$ is the ionization potential.
For the first process in Eq.(\ref{14x}), $I = 24.6$ eV while in the second process, $I = 54.4$ eV. We denote
the corresponding cross-sections as $\sigma_I^{a}$ and $\sigma_I^b$ respectively.

Opposing ionization are the $e'$ capture processes.
The relevant processes for $He'$ are:
\begin{eqnarray}
e ' + {He'}^{+}  &\rightarrow & {He'}^{0}  + \gamma '  
\nonumber \\
e ' + {He'}^{2+}  &\rightarrow & {He'}^{+}  + \gamma '  
\ .
\label{go}
\end{eqnarray}
The cross-section for the capture processes can be approximated by a 
modified Kramers formula\cite{pratt}:
\begin{eqnarray}
\sigma_C = \sum_n {8\pi \over 3\sqrt{3}} {\alpha^5 \over n^3} {Z_{eff}^4 \over E_{e'} E_{\gamma'}}
\label{cap}
\end{eqnarray}
where $E_{\gamma'} = E_{e'} + {Z^2_{eff}\alpha^2 m_e \over 2n^2}$.
For the applications to $He', H'$ and also $Fe'$ ions
that we will consider, $Z_{eff} = (Z_C + Z_I)/2$, where $Z_C$ is the charge
of the nuclei and $Z_I$ is the ionic charge before $e'$ capture\cite{pratt}.
Thus, $Z_{eff} \approx 1.5$ for the first process in Eq.(\ref{go}), and $Z_{eff} = 2$ for the
second process. We denote the corresponding cross-sections as $\sigma_C^{a}$ and $\sigma_C^{b}$
respectively.

The above processes dictate the number density of ${He'}^{2+}$ via:
\begin{eqnarray}
{dn_{{He'}^{2+}} \over dt} = 
n_{e'} n_{{He'}^{+}}
\langle
\sigma_I^b v_{e'} \rangle
- 
n_{e'} n_{{He'}^{2+}}
\langle \sigma_C^b  
v_{e'} \rangle
\ 
\end{eqnarray}
where the brackets $\langle ... \rangle$
indicate the average over the $e'$ velocity distribution
taken as a Maxwell-Boltzman distribution: 
\begin{eqnarray}
\langle \sigma^b_I v_{e'} \rangle &\equiv & 2 \sqrt{ {2 \over m_e \pi} } \left(
{1 \over T}\right)^{3/2} \ \int_I^{\infty}
\sigma^b_I
\ e^{-E_{e'}/T} 
\ E_{e'}\ dE_{e'}
\nonumber \\
\langle \sigma^b_C v_{e'} \rangle &\equiv & 2 \sqrt{{2 \over m_e \pi }} \left(
{1 \over T}\right)^{3/2} \ \int_0^{\infty}
\sigma^b_C
\ e^{-E_{e'}/T} 
\ E_{e'}\ dE_{e'}
\ .
\end{eqnarray}
In a steady state situation, we have $dn_{{He'}^{2+}}/dt = 0$ and thus
\begin{eqnarray}
R_2^{He'} \equiv {n_{{He'}^{2+}}
\over n_{{He'}^{+}}} = 
{
\langle \sigma_I^b v_{e'}\rangle 
\over 
\langle
\sigma_C^b v_{e'} \rangle 
}
\ .
\end{eqnarray}
Similarly for the processes affecting $dn_{{He'}^0}/dt$, and for the 
corresponding process for mirror hydrogen:
\begin{eqnarray}
R_1^{He'} \equiv {n_{{He'}^{+}}
\over n_{{He'}^{0}}} &=& 
{
\langle \sigma_I^a   
v_{e'} \rangle
\over 
\langle \sigma_C^a
v_{e'}\rangle
}
\nonumber \\
R_1^{H'} \equiv {n_{{H'}^{+}}
\over n_{{H'}^{0}}} &=& 
{ 
\langle \sigma_I v_{e'} \rangle
\over 
\langle \sigma_C v_{e'}\rangle}
\end{eqnarray}
where $\sigma_I$ and $\sigma_C$ are the relevant cross-sections for the mirror hydrogen 
process. The $H'$ ionization cross-section is given by Eq.(\ref{ion}) with $I = 13.6$ eV,
while the capture cross-section is given by Eq.(\ref{cap}) with $Z_{eff} = 1$.

With these definitions we can determine the number density of each component as a function of one of them,
which we choose to be $n_{He'} = n_{{He'}^0} + n_{{He'}^+} + n_{{He'}^{2+}}$:
\begin{eqnarray}
n_{{He'}^{2+}} &=& \left( {R_1^{He'} R_2^{He'} \over 1 + R_1^{He'} + R_1^{He'} R_2^{He'}}\right) \ n_{He'} \nonumber \\
n_{{He'}^{+}} &=& \left( {R_1^{He'}  \over 1 + R_1^{He'} + R_1^{He'} R_2^{He'}} \right) \ n_{He'} \nonumber \\
n_{{He'}^{0}} &=& n_{He'} - n_{{He'}^{+}} - n_{{He'}^{2+}} \nonumber \\
n_{{H'}^{+}} &=& \left( {R_1^{H'} \over 1 + R_1^{H'}} \right) \ f n_{He'} \nonumber \\
n_{{H'}^{0}} &=& f n_{He'} - n_{{H'}^{+}} \nonumber \\
n_{e'} &=& 2n_{{He'}^{2+}} + n_{{He'}^{+}} + n_{{H'}^{+}} \nonumber \\
n_{T} &=& (1 + f)n_{{He'}} + n_{e'} 
\end{eqnarray}
where $f \equiv n_{H'}/n_{He'}$ and $n_T$ is the total particle number density. 
The fraction, $f$, can be related to the $He'$ mass fraction:
\begin{eqnarray}
\xi_{He'}
= {1 \over 1 + f/4}
\ .
\end{eqnarray}
Unless otherwise stated, we take $f=0.4$ in our numerical work (which, as we already discussed at the beginning
of the previous section, is suggested from early Universe cosmology).
The quantities, $R_{1,2}^{He'}$ and $R_1^{H'}$  depend only on the temperature. It is straightforward
to compute the $He'$ ionization fractions, 
$F^{He'}_0 \equiv n_{{He'}^{0}}/n_{He'}$, 
$F^{He'}_1 \equiv n_{{He'}^{+}}/n_{He'}$, 
$F^{He'}_2 \equiv n_{{He'}^{2+}}/n_{He'}$ and
also the $H'$ ionization fractions, 
$F^{H'}_0 \equiv n_{{H'}^{0}}/n_{H'}$, 
$F^{H'}_1 \equiv n_{{H'}^{+}}/n_{H'}$. 
We show these results in figure 2 for $He'$ and figure 3 for $H'$.
Figure 2 indicates that $He'$ is nearly fully ionized for $T \stackrel{>}{\sim}$ 10 eV. 
This is substantially below the, $I = 54.4$ eV ionization energy of ${He'}^{+}$, which occurs because
the capture cross-section is several orders of magnitude smaller than the ionization cross-section.
Qualitatively similar results arise also for $H'$.

\centerline{\epsfig{file=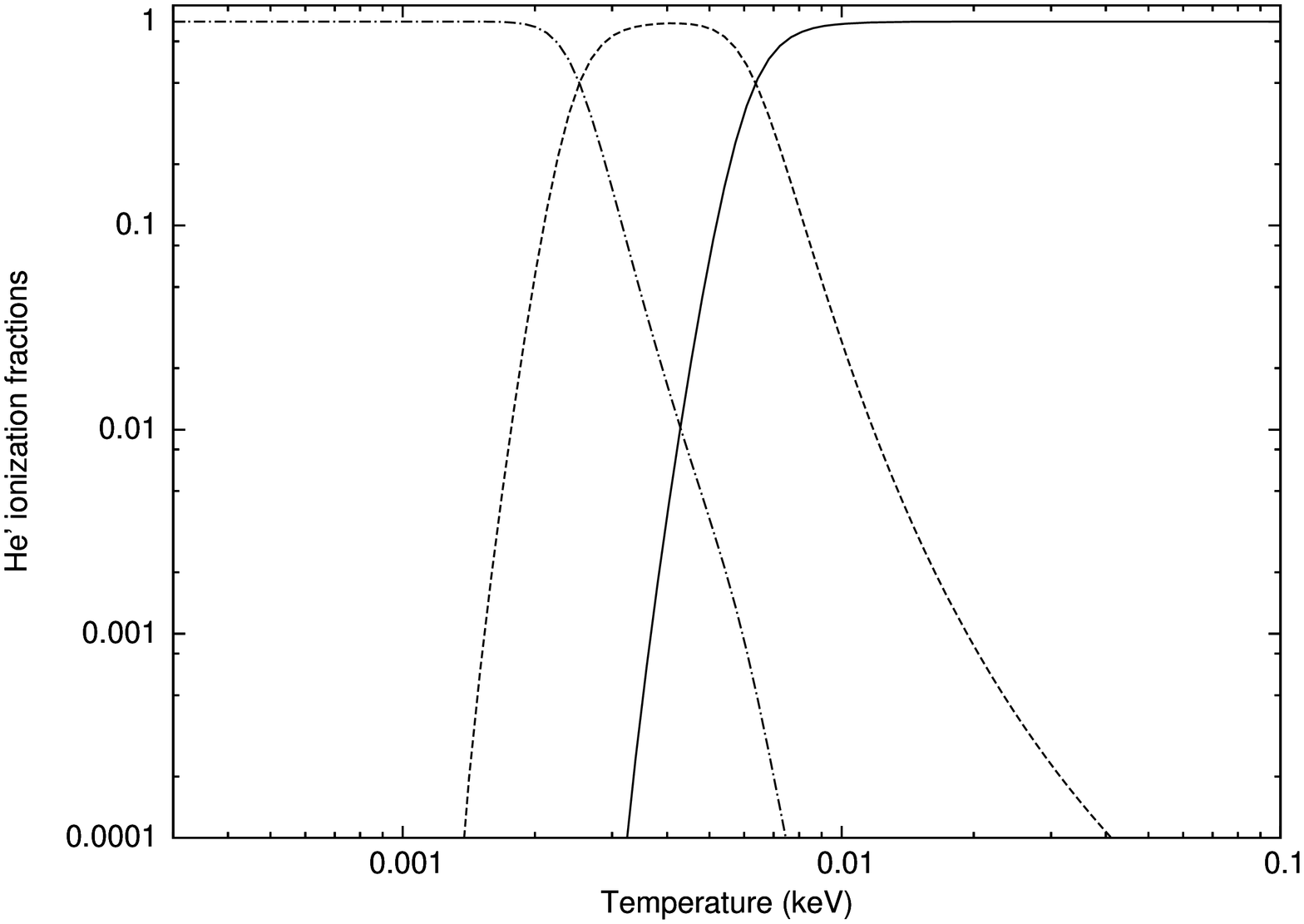, angle=270, width=12.8cm}}
\vskip 0.4cm
\noindent
{\small Figure 2: The $He'$ ionization fractions,
as a function of the local halo temperature, $T$.
Shown are 
$F^{He'}_0 \equiv n_{{He'}^{0}}/n_{He'}$ (dashed-dotted line), 
$F^{He'}_1 \equiv n_{{He'}^{+}}/n_{He'}$ (dashed line) 
and $F^{He'}_2 \equiv n_{{He'}^{2+}}/n_{He'}$ (solid line). 
}

\vskip 0.2cm
\centerline{\epsfig{file=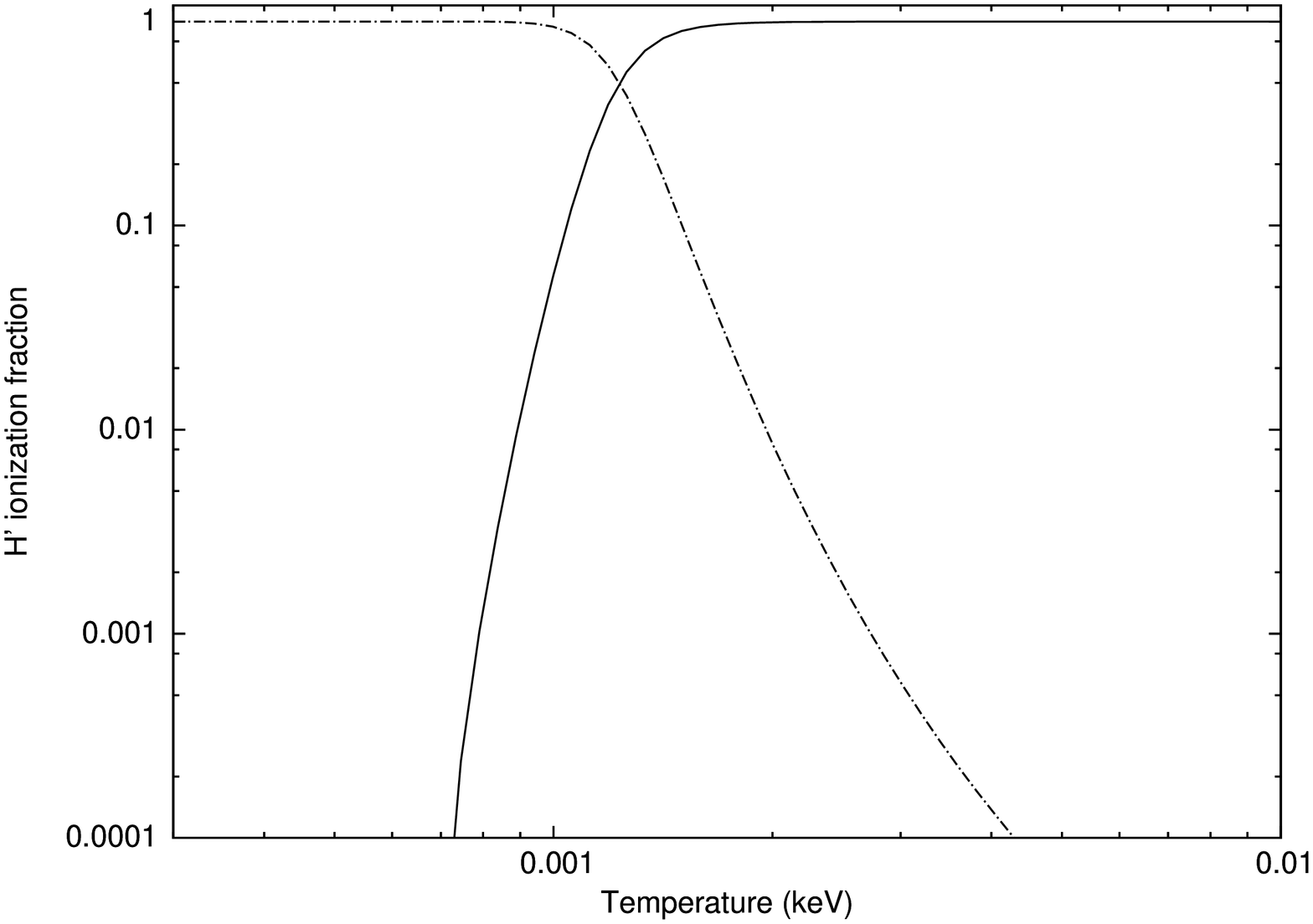, angle=270, width=12.8cm}}
\vskip 0.4cm
\noindent
{\small Figure 3: The $H'$ ionization fractions,
as a function of the local halo temperature, $T$.
Shown are 
$F^{H'}_0 \equiv n_{{H'}^{0}}/n_{H'}$ (dashed-dotted line) 
and $F^{H'}_1 \equiv n_{{H'}^{+}}/n_{H'}$ (solid line). 
}

\vskip 1.2cm

In addition to the pure $H', He'$ halo, we will consider a small metal component, 
which we take as $Fe'$ with
total number density, $n_{Fe'}$ [the $Fe'$ component was presumably formed in mirror
stars at an early epoch, see earlier discussion above Eq.(\ref{mbar})].
We denote the number density of completely ionized $Fe'$ as $n_{{Fe'}^{**}}$ and  $Fe'$ with 1 K-shell $e'$ 
as $n_{{Fe'}^*}$.
The ionization energy of the bound mirror electron in ${Fe'}^{*}$ is 9.3 keV
and if both K-shell mirror electrons are present, the binding energy is 8.8 keV\cite{sugar}.
In figure 4 we show
the 
computed ionization fractions, $F^{Fe'}_1 \equiv n_{{Fe'}^*}/n_{Fe'}$ and
$F^{Fe'}_2 \equiv n_{{Fe'}^{**}}/n_{Fe'}$ versus temperature.
Figure 4 indicates that $Fe'$ is nearly fully ionized until the temperature drops below around $20$ keV 
i.e. somewhat above the ionization energy of the K-shell bound mirror electrons.
[This occurs because the ${Fe'}^*$ capture cross-section is somewhat larger than the ionization cross-section.]
For temperatures below around 2 keV, greater than 99 percent of the $Fe'$ ions typically has both atomic K-shell states filled.

\centerline{\epsfig{file=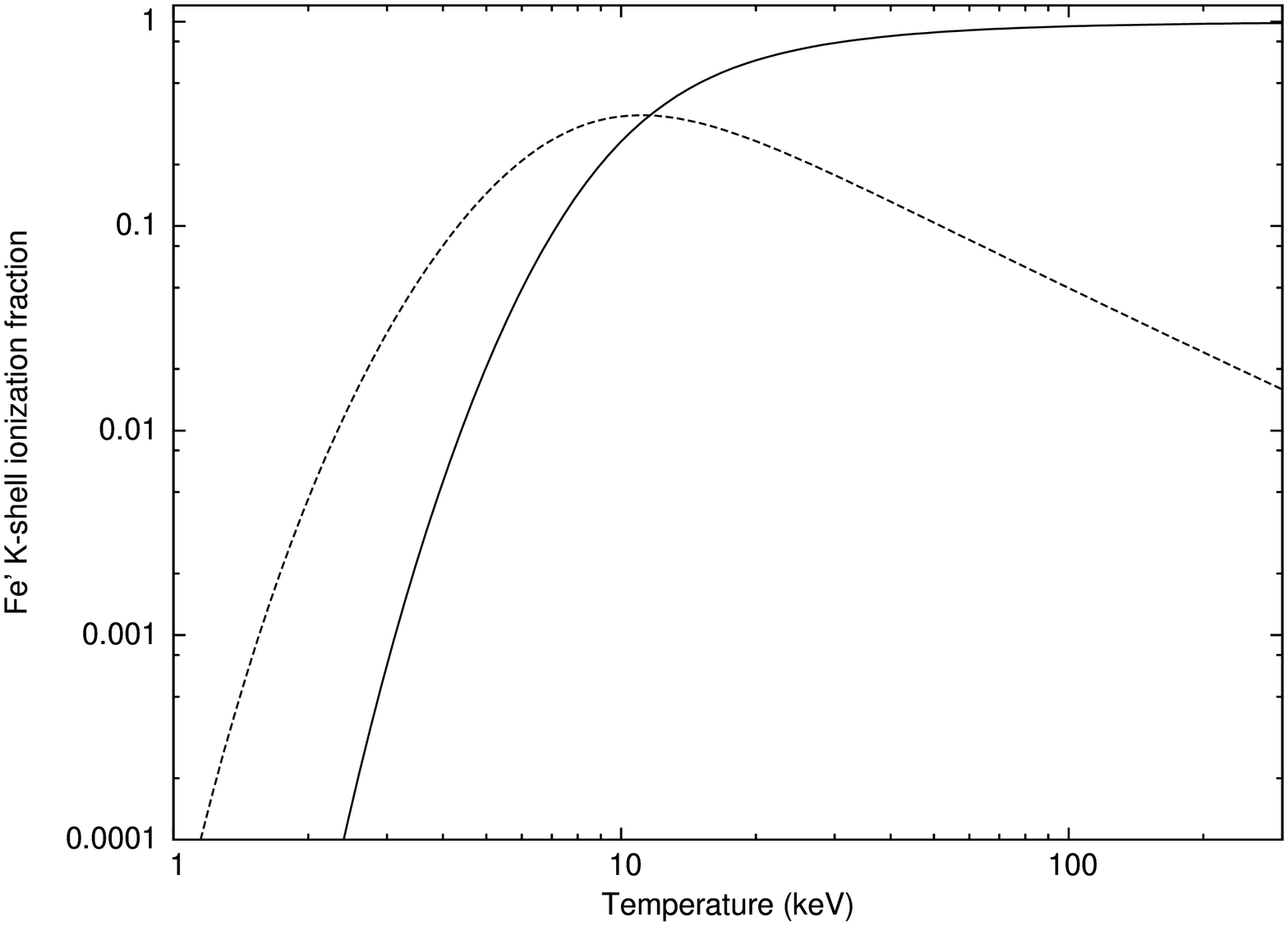, angle=270, width=12.8cm}}
\vskip 0.4cm
\noindent
{\small Figure 4: The $Fe'$ ionization fractions,
$F^{Fe'}_1 \equiv n_{{Fe'}^{*}}/n_{Fe'}$ (dashed line), $F^{He'}_2 \equiv n_{{Fe'}^{**}}/n_{Fe'}$ (solid line)
as a function of the local halo temperature, $T$.
}

\vskip 2.9cm

\newpage

\section{Halo energetics and scaling relations}

Studies of spiral galaxies have found that the baryonic parameters, $m_D, r_D, L_r$
satisfy two approximate relations. One of these relations was given in the
previous section, Eq.(\ref{rd}), the other 
relates the r*-band luminosity, $L_r$ to $m_D$ (see e.g. \cite{sal3} and references there-in):
\begin{eqnarray}
{m_D \over 10^{11} m_\odot}
\approx
\left( {L_r \over L_r^{MW}} \right)^{1.3}
\label{mdrel}
\end{eqnarray}
where $L_r^{MW} \approx 2\times 10^{10} L_\odot$ is the r*-band luminosity of the Milky Way.
With these relations, the baryonic parameters of spirals are (roughly) specified by a single
parameter which can be taken as one of $m_D, r_D$ or $L_r$.
If we assume that the dark matter is distributed via the Burkert profile, Eq.(\ref{1}),
then we have two further parameters, $\rho_0$ and $r_0$. Our aim is to 
derive constraints on these two dark matter parameters from dynamical considerations.
The derived constraints can then be compared with `empirical' relations
derived from observations of rotation curves in spiral 
galaxies\cite{PSS,kf,sal2,donato,sal3,dsds,core}:
\begin{eqnarray}
log\left( {\rho_0 r_0 \over m_\odot {\rm pc}^{-2}} \right) &\simeq  & 2.2 \pm 0.25 \nonumber \\
{L_r \over 1.2\times 10^{10} L_\odot} &\simeq & 
{\left({m_h \over 3 \times 10^{11} m_\odot}\right)^{2.65} \over 
1 + \left( {m_h \over 3\times 10^{11} m_\odot}\right)^{2.00} 
}
\nonumber \\
log \left( {r_0 \over {\rm kpc}} \right) &\simeq & 0.66 + 0.58 \ log \left( {m_h \over 10^{11} m_\odot}  
\right)
\label{rel1}
\end{eqnarray}
where 
$m_h$ is the halo mass.
Spiral galaxies typically have halo's in the mass range,
$10^{11} m_\odot \stackrel{<}{\sim} m_h \stackrel{<}{\sim} 10^{13} m_\odot$ and
baryonic mass in the range,
$10^{9} m_\odot \stackrel{<}{\sim} m_D \stackrel{<}{\sim} 10^{12} m_\odot$.

As discussed in refs.\cite{sph,r69} and reviewed in section 2, 
the energy lost in the halo due to dissipative processes might be
replaced by supernova energy transported to the halo via mirror photons\footnote{
One can check that the energy transport due to other processes, such as conduction,
is negligible in comparison to radiation. This is, of course, due to the fact that the 
mirror photons have a much longer scattering length than mirror electrons.}.
For this idea to work out, the halo must evolve to a state such that
the energy being absorbed in each volume element must equate to the energy being
radiated from the same volume element.
Thus, we have a dynamical condition:
\begin{eqnarray}
{d^2 E_{in} \over dt dV}= 
{d^2 E_{out} \over dt dV} 
\ .
\label{meal8}
\end{eqnarray}
In the following, we derive approximate formula for the left and right-hand sides 
of the above equation. Subsequently we solve the derived equations numerically. We will show 
that the above dynamical condition can be approximately satisfied with the  Burkert dark matter profile
provided that $\rho_0$ and $r_0$ satisfy
certain relations.
These relations are then compared with the `empirical' relations obtained from galactic
rotation curve (and other) data, discussed above.

The region around a single supernova will be a source of a huge flux of mirror photons
with total integrated luminosity of around $10^{53}$ erg, provided photon - mirror photon
kinetic mixing of strength, $\epsilon \sim 10^{-9}$ exists.
We define the total (time) average mirror photon luminosity 
due to ordinary core collapse supernovae by
\begin{eqnarray}
L'_{SN}  \equiv R_{SN} f_{SN} \langle E_{SN} \rangle
\label{fsn}
\end{eqnarray}
where $R_{SN}$ is the galactic supernova rate, $f_{SN}$ is the fraction of supernovae 
energy emitted by mirror particles, 
$e', \bar e', \gamma'$, 
and $\langle E_{SN} \rangle$ is the
average total energy emitted per supernova. 
The quantity $L'_{SN}$
obviously depends on the particular 
galaxy concerned.   For the Milky Way galaxy, we have:
\begin{eqnarray}
{L'}_{SN}^{MW} \approx 
\left({R_{SN}^{MW} \over 0.03 \ {\rm yr}^{-1}} \right) 
\left( {f_{SN} \over 0.5}\right)
\left({\langle E_{SN} \rangle \over 3\times 10^{53}\ {\rm erg}} \right) 
\ 1.4\times 10^{44} \ {\rm erg/s}
\ .
\end{eqnarray}
To proceed further we will need to parameterize 
the $\gamma'$ supernova energy spectrum averaged over all ordinary supernova.
We assume that the peak of this (averaged) $\gamma'$ energy spectrum occurs at energies 
somewhat greater than the K-shell $e'$ atomic binding
energy of $Fe'$, $\sim 9$ keV.
In this case only the low energy part of the spectrum can heat the halo.
We parameterize this energy spectrum via a power law:
\begin{eqnarray}
E_{\gamma'} {dN_{\gamma'} \over dE_{\gamma'}} &=& \left( {1 + c_1 \over E_c}\right)\left(
{E_{\gamma'}\over E_c}\right)^{c_1} \ f_{SN} E_{SN}\nonumber \\
&\equiv & \kappa \left(E_{\gamma'}\right)^{c_1}
\ .
\label{m3}
\end{eqnarray}
This spectrum has been normalized such that 
\begin{eqnarray}
\int^{E_c}_0 E_{\gamma'} {dN_{\gamma'} \over dE_{\gamma'}} \ dE_{\gamma'} = f_{SN} E_{SN}
\ .
\end{eqnarray}
We will consider $1 \le c_1 \le 3$ in our numerical work ($c_1 = 2$ corresponds to a thermal spectrum).
Although the spectrum would not be expected to be a power law for energies sufficiently high, such details will be
unimportant since the halo is optically thin at energies $E_\gamma' \stackrel{>}{\sim} 30$ keV.

Observations indicate that the supernova rate scales with galactic 
B-band luminosity, $L_B$, 
via $R_{SN} \propto \left(L_B\right)^{0.73}$,  
with an uncertainty in the exponent around 0.1\cite{sn}.
With the above definitions, we find for $f_{SN}
\approx 0.5$ (i.e. for $\epsilon \sim 10^{-9}$),
\begin{eqnarray}
\kappa\ R_{SN} \approx {1 + c_1 \over (E_c)^{1+c_1}}
\left( {L_B \over L_B^{MW}}\right)^{0.73}
\ {L'}_{SN}^{MW}
\label{kap2}
\end{eqnarray}
where $L_B^{MW} \approx 2 \times 10^{10} L_\odot$ is the reference B-band luminosity for the Milk Way. 
Although there is substantial uncertainty in ${L'}_{SN}^{MW}$, possibly as large 
as an order of magnitude, the galactic scaling behaviour of $\kappa\ R_{SN}$ 
should be more certain.  

\centerline{\epsfig{file=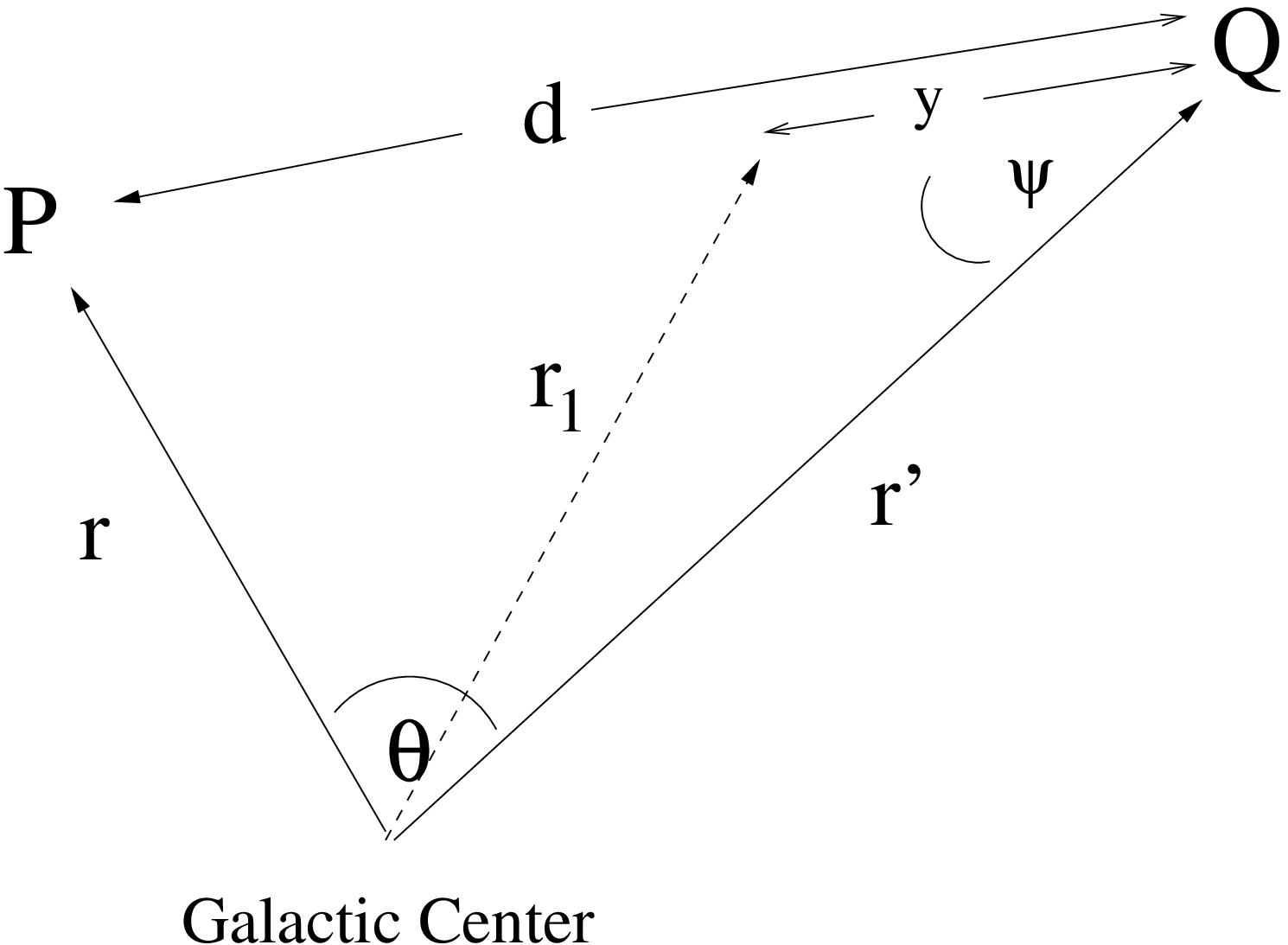, angle=0, width=7.0cm}}
\vskip 0.5cm
\noindent
{\small Figure 5: The geometry. Mirror photons travel a distance $d$ from a supernova source at point $Q$, to heat the halo at a point $P$.
}
\vskip 1.2cm

The energy spectrum of $\gamma'$ from a single supernova source is given in Eq.(\ref{m3}).
The supernovae are distributed throughout the disk. We wish to figure out the (time) average
$\gamma'$ flux at a point, $P$, a distance $r$
from the galactic center, from all supernova sources.
We assume that the supernova distribution traces the baryonic mass density [Eq.(\ref{12z})]. It then
follows that
the contribution to the $\gamma'$ energy flux at $P$ originating from a 
volume element, $dV' = 2\pi r'^2 \ d\cos\theta dr' $, at a point $Q$ is
\begin{eqnarray}
{d^2F(r) \over dE_{\gamma'}dV'} 
= {\kappa \left(E_{\gamma'}\right)^{c_1} R_{SN} \ e^{-\tau } \over 4\pi {\rm d}^2} {\rho_D \over m_D} 
\label{kap}
\end{eqnarray}
where $\kappa$ is defined in Eq.(\ref{m3}), ${\rm d} = \sqrt{ r^2 + r'^2 - 2rr' \cos\theta}$ 
is the distance of the source $Q$ to the point $P$ and
$\tau$ is the optical depth along that path.
Summing over all contributions we find that the 
total differential energy flux is given by:
\begin{eqnarray}
{dF(r) \over dE_{\gamma'}} = {\kappa \left( E_{\gamma'}\right)^{c_1} R_{SN} \over m_D} 
\int^{\infty}_0 \int^1_{-1} \ {\rho_D \ e^{-\tau} \ r'^2 \over 2{\rm d}^2}\ d\cos\theta dr'
\ .
\label{33a}
\end{eqnarray}
The optical depth $\tau$ is given by
\begin{eqnarray}
\tau = \int_0^{\rm d} \sum_i n_i(r_1) \sigma_i dy
\ .
\label{str}
\end{eqnarray}
The relevant geometry is shown in figure 5, and we have
\begin{eqnarray}
r_1 &=& \sqrt{y^2 + r'^2 - 2r'y\cos\psi} \nonumber \\
\cos \psi &=& {{\rm d}^2 + r'^2 - r^2 \over 2r'{\rm d}} \ .
\end{eqnarray}

The flux of supernova $\gamma'$ at a particular point, $P$, will deposit an energy per unit volume per unit time
of: 
\begin{eqnarray}
{d^3 E_{in} \over dE_{\gamma'}dt dV}= {dF \over dE_{\gamma'}}\sum_i n_i (r) \sigma_i 
\ .
\label{meal}
\end{eqnarray}

Thus to calculate the heating at a particular point in the halo we need to determine the
cross-section 
and number densities of the various components in the halo [denumerated by $i$, in Eq.(\ref{str}) and Eq.(\ref{meal})].
The supernova $\gamma'$ are assumed to have relatively high energies, so that the optical depth is
dominated by scattering of $\gamma'$ off bound atomic $e'$. 
This is possible because
heavy elements, such as $Si',Fe'$, are not completely ionized but have their atomic inner shells filled.
If the flux of supernova $\gamma'$ are falling below 10 keV, then the interactions with K-shell $Fe'$ mirror electrons
are likely to be the most important (see discussion in section 2).
The total photoelectric cross-section 
is given approximately by Eq.(\ref{pe4}) and  
the $Fe'$ number density is given by 
$n_{Fe'} = n_{He'} \left(1 + {f \over 4}\right) \left({m_{He} \over m_{Fe}}\right) \left({\xi_{Fe'} \over 1 - \xi_{Fe'}}\right)
$.
[Also needed are the $Fe'$ ionization fractions which can be computed as per section 4.]

Having discussed the heating rate at the point $P$ in the halo, we now turn our attention to the cooling rate at the
same point.
The cooling rate is expected to have contributions from three
sources: thermal bremsstrahlung, line emission and recombination.
We first consider the bremsstrahlung 
component and comment on the line emission and recombination
contributions in a moment.

The rate at which bremsstrahlung energy is radiated per unit volume, per unit time 
is\cite{bookastro}:
\begin{eqnarray}
{d^2 W \over dt dV} = {16\alpha^3 \over 3m_e}\left({
2\pi T \over 3m_e}\right)^{1/2} 
\ \sum_j
\left[ Z_j^2 n_j n_{e'} \bar g_B  \right]
\label{john}
\end{eqnarray}
where the index $j$ runs 
over the mirror ions in the plasma (of charge $Z_j$)  and
$\bar g_B$ is the frequency average of the velocity averaged
Gaunt factor for free-free emission. 
We take $\bar g_B = 1.2$, which, as reviewed in ref.\cite{bookastro}, should be 
accurate to within about 20\%. 

In principle the energy radiated at a point, $P$, can have important contributions from
line emission and recombination in addition to bremsstrahlung.  
Such processes would depend on the detailed chemical composition of the halo. 
However these processes may not be as important as naive first thoughts suggest.
The energy of the radiated mirror photons
from line emission and recombination is typically close to the halo temperature, $T$. But the halo
is generally expected to be optically thick to mirror photons
of these energies (rough estimates indicate). Thus, the capacity of these 
processes to directly cool the halo is 
expected to be greatly diminished. 
Bremsstrahlung, on the other hand, generally produces mirror photons of lower energy.
The energy spectrum of bremsstrahlung is flat for $E_{\gamma'} \ll T$.
and reduces towards zero for increasing $T$ like $\sim exp(-E_{\gamma'}/T)$
(see e.g. \cite{bookastro}).
We therefore expect that the bremsstrahlung
process will cool the halo much more efficiently.
As a rough approximation, we set ${d^2 E_{out} \over dt dV} = \epsilon_f 
{d^2W \over dt dV}$, where
$\epsilon_f$ is an efficiency factor which we set to unity in our numerical work.
If the bremsstrahlung process is the dominant cooling
mechanism then ${dE_{out}\over dt dV} \propto \sqrt{T}$, that is, an increasing
function of $T$.
This might explain why the system evolves until $d E_{in} = dE_{out}$.
If a region had $dE_{in} > dE_{out}$ then this will make $T$ higher in that region which increases
also $dE_{out}$ until $dE_{in} = dE_{out}$. Similarly if 
$dE_{in} < dE_{out}$ then this will make $T$ smaller which decreases
$dE_{out}$ until $dE_{in} = dE_{out}$. 
It therefore seems plausible that the system will evolve
until the dynamical condition, Eq.(\ref{meal8}) is satisfied everywhere\footnote{
Naturally the system is a complicated one and other feedback mechanisms can also be 
important. For instance, a mismatch of $dE_{in}$ and $dE_{out}$ can also cause
expansion or contraction of the halo, which in turn can affect the ordinary star formation rate and
thereby readjust $dE_{in}$.
Such a feedback mechanism may also help regulate the star formation rate as suggested by 
observations [see ref.\cite{tut} and references there-in for relevant discussions].}.

We are now ready to start solving the equations.
The philosophy is that if the dark matter density can be parameterized
by the form given in Eq.(\ref{1}) then $\rho_0$ and $r_0$
can be determined by demanding that ${d^2 E_{in} \over dt dV} \simeq 
{d^2 E_{out} \over dt dV}$ for each
volume element.
To quantify how well $dE_{in}$ and $dE_{out}$ match, we introduce the quantity, $\Delta$:
\begin{eqnarray}
\Delta (r_0, \rho_0) \equiv {1 \over 10 r_D} \int^{11r_D}_{r_D}
\left| 1 - {{d^2 E_{in} \over dt dV} \over {d^2 E_{out} \over dt dV}}\right| \ dr
\ .
\label{delta}
\end{eqnarray}
The quantity $\Delta$ can be computed numerically via a fortran code.
The adopted procedure is to input the dark matter density profile and the baryonic density and work out the temperature
profile using the hydrostatic equilibrium condition as described in section 3. Once
the temperature profile is known, one can work out the ionization state of the halo
via the equations described in section 4. 
Armed with this information, one can proceed to work out Eqs.(\ref{33a},\ref{meal}) and Eq.(\ref{john}),
the latter we equate to ${d^2 E_{out} \over dt dV}$.
The end result is the trivial integration, Eq.(\ref{delta}), to obtain $\Delta$ as a function of the input parameters,
$\rho_0, \ r_0$ as well as baryonic parameter, $m_D$ [$r_D$ and $L$ are obtained from Eq.(\ref{rd}) and Eq.(\ref{mdrel})].

We start by examining how compatible the Burkert 
dark matter profile is with the ${d^2 E_{in} \over dt dV} = {d^2 E_{out} \over dt dV}$ condition.
Define $\Delta^a_{min} (r_0)$ as the quantity $\Delta (r_0, \rho_0)$ minimized
with respect to variations in $\rho_0$.
Similarly,
$\Delta^b_{min} (\rho_0)$ is defined by minimizing  $\Delta (r_0, \rho_0)$ 
with respect to variations in $r_0$.
In figure 6a we plot $\Delta^a_{min} (r_0)$ versus $r_0$ 
and in figure 6b we plot $\Delta^b_{min} (\rho_0)$ versus $\rho_0$.
These figures are for an
example point with $m_D = 10^{11} m_\odot$ and reference parameters,
$f\equiv n_{H'}/n_{He'} = 0.4$, $\xi_{Fe'} = 0.02$, ${L'}_{SN}^{MW} = 2\times 10^{45}$ erg/s, $E_c = 50$ keV. 

\centerline{\epsfig{file=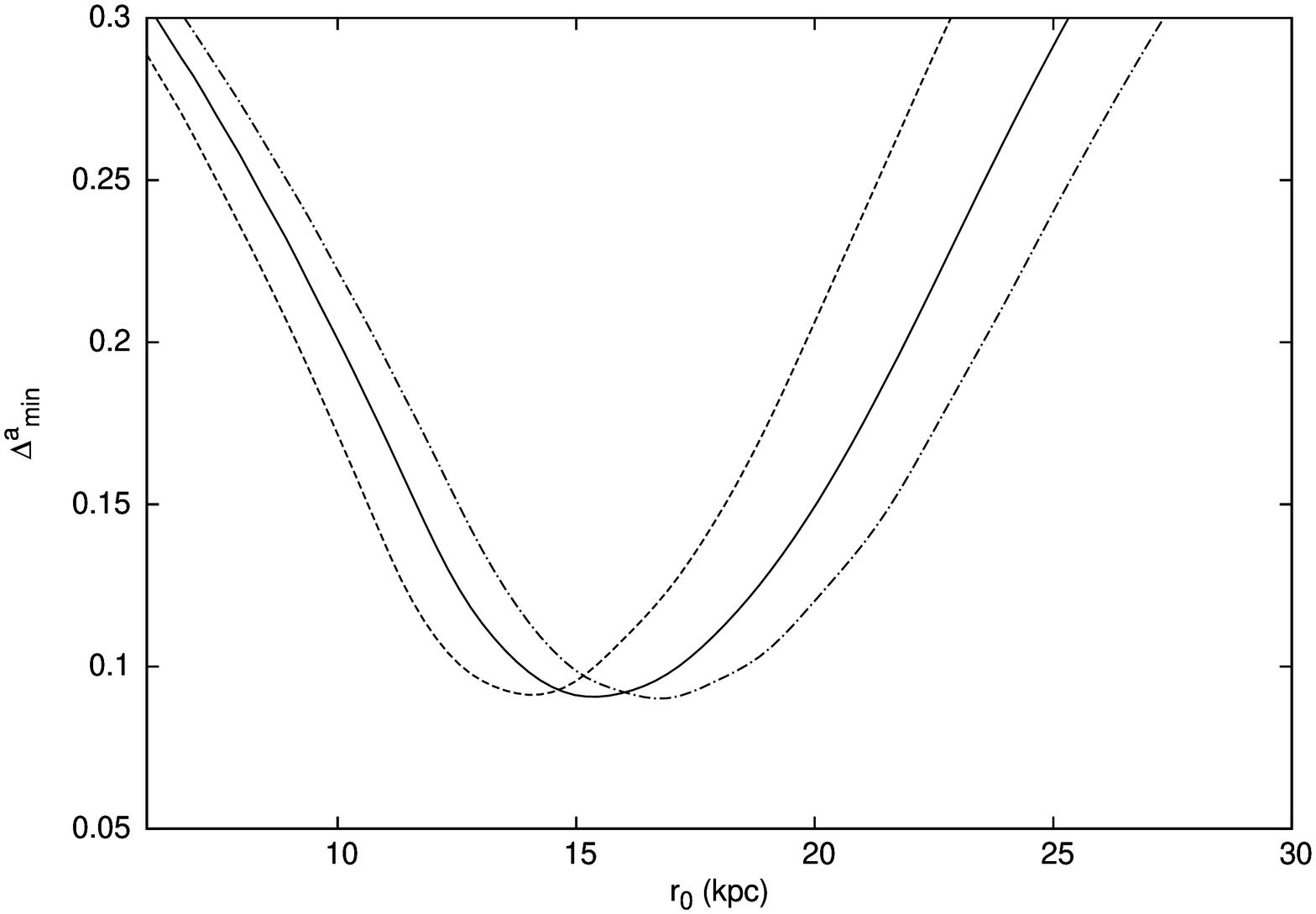, angle=270, width=12.1cm}}
\vskip 0.2cm
\noindent
{\small Figure 6a: 
$\Delta^a_{min}$ (defined in text) versus the core radius, $r_0$
for $m_D = 10^{11} \ m_\odot$.
Plotted are various values of $c_1$ which parameterize the hardness
of the mirror photon supernova spectrum: $c_1 = 1$ (dashed line), 
$c_1 = 2$ (solid line) and $c_1 = 3$ (dashed-dotted line).
}
\vskip 0.2cm
\centerline{\epsfig{file=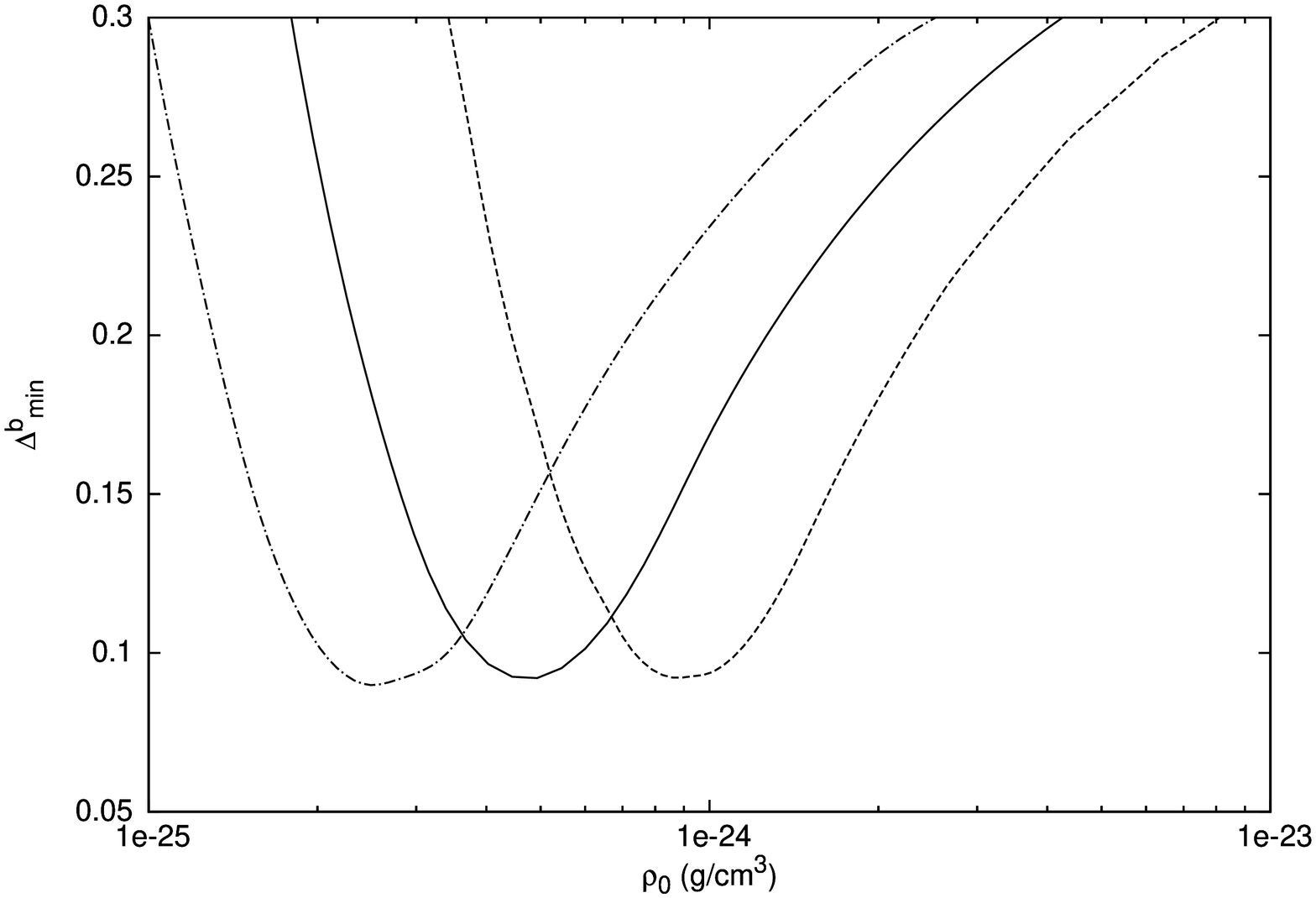, angle=270, width=12.8cm}}
\vskip 0.3cm
\noindent
{\small Figure 6b: 
Same as for figure 6a, except $\Delta^b_{min}$ is plotted versus $\rho_0$.
}
\vskip 0.2cm
\centerline{\epsfig{file=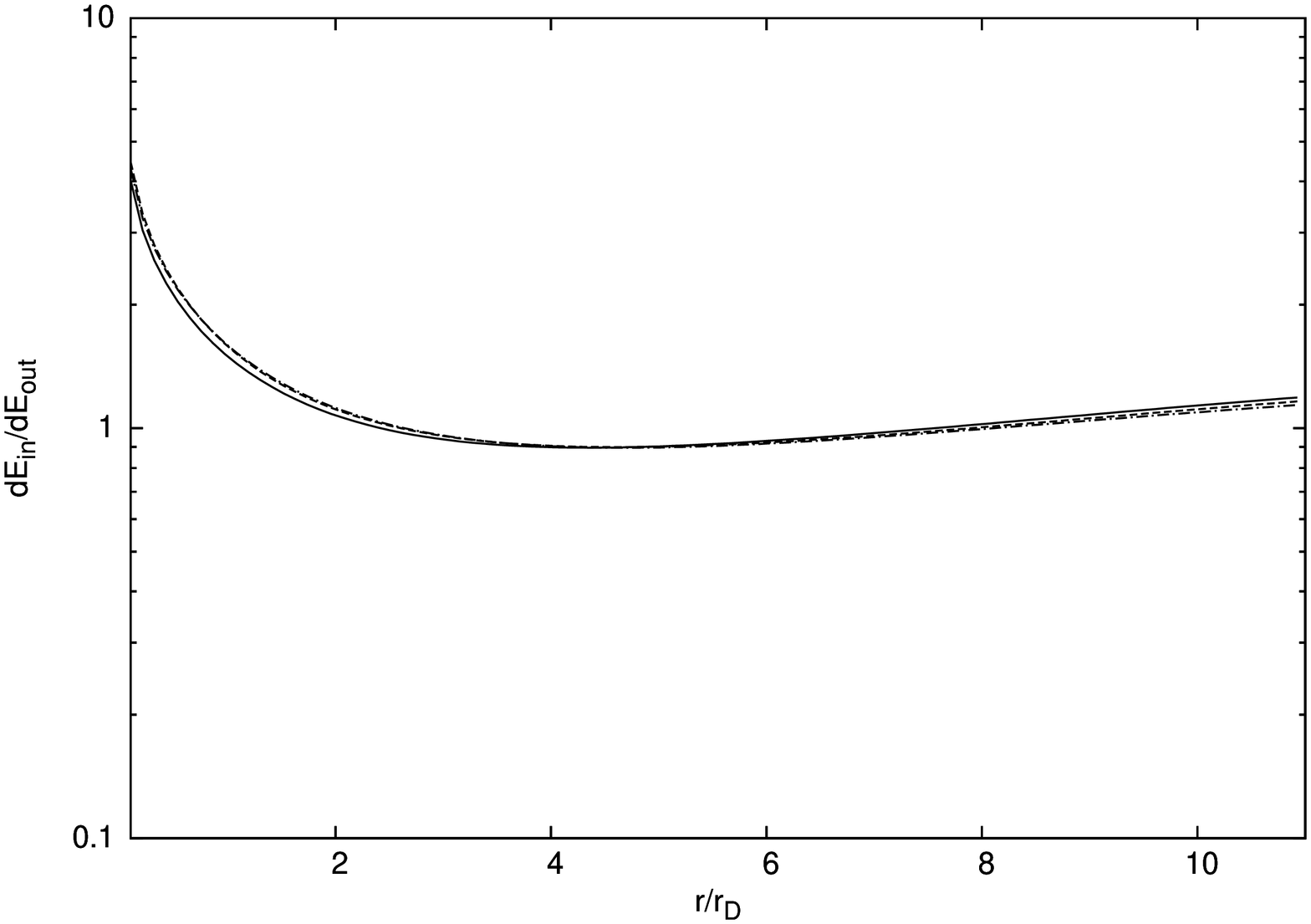, angle=270, width=12.8cm}}
\vskip 0.3cm
\noindent
{\small Figure 6c: 
${d^2 E_{in} \over dt dV}/{d^2 E_{out} \over dt dV}$ versus  $r/r_D$ 
for the same example as figure 6a,b. 
The value of $r_0, \rho_0$ taken are the one's which minimize $\Delta$. 
Three curves (almost indistinguishable) correspond to the same three $c_1$ 
values of figure 6a,b.
}
\vskip 1.0cm
The figures show that $\Delta$
has a minimum value of around $0.1$.  That is,
${d^2 E_{in} \over dt dV} = {d^2 E_{out} \over dt dV}$ to within around 10\%. 
In figure 6c, we plot 
${d^2 E_{in} \over dt dV}/{d^2 E_{out} \over dt dV}$ 
(evaluated at the $\rho_0,\ r_0$ values which minimize $\Delta$) 
versus galactic radius $r/r_D$. This is done for the same example
as figure 6a,b. 
This figure demonstrates that the Burkert profile does a very good job at minimizing $\Delta$
with only small deviations near the galactic center, 
which are unlikely to be important since ordinary baryons dominate the mass density there (see e.g.\cite{core}
and references there-in).
We have found that
varying $m_D$ and variations in the other parameters give similar results.
This demonstrates that the Burkert profile is (roughly) compatible with the condition, Eq.(\ref{meal8}).
Furthermore, for each value of $m_D$ we can estimate the values of $r_0$ and $\rho_0$
by minimizing $\Delta$, as we have just done for the particular example with $m_D = 10^{11} m_\odot$.
This leads to relations connecting the baryonic parameters ($m_D, \ r_D, \ L$)
with the dark matter parameters, $\rho_0, r_0$.
Although we expect to derive only two independent relations, we explore our results by 
considering three plots.
In figure 7a we plot $L$ versus $r_0$,   
in figure 7b we plot $\rho_0$ versus $m_D$ and 
in figure 7c we 
plot $\rho_0 r_0$ versus $L$. 
In each case we consider various values of $c_1$ which, recall, parameterizes 
the hardness of the supernova $\gamma'$ spectrum.
Other parameters fixed are
$f \equiv n_{H'}/n_{He'} = 0.4$, $\xi_{Fe'} = 0.02$. The quantity
$\kappa \ R_{SN}$ 
is given in Eq.(\ref{kap2}) with $E_c = 50$ keV and 
${L'}_{SN}^{MW} = 2\times 10^{45} \ {\rm erg/s}$.

\centerline{\epsfig{file=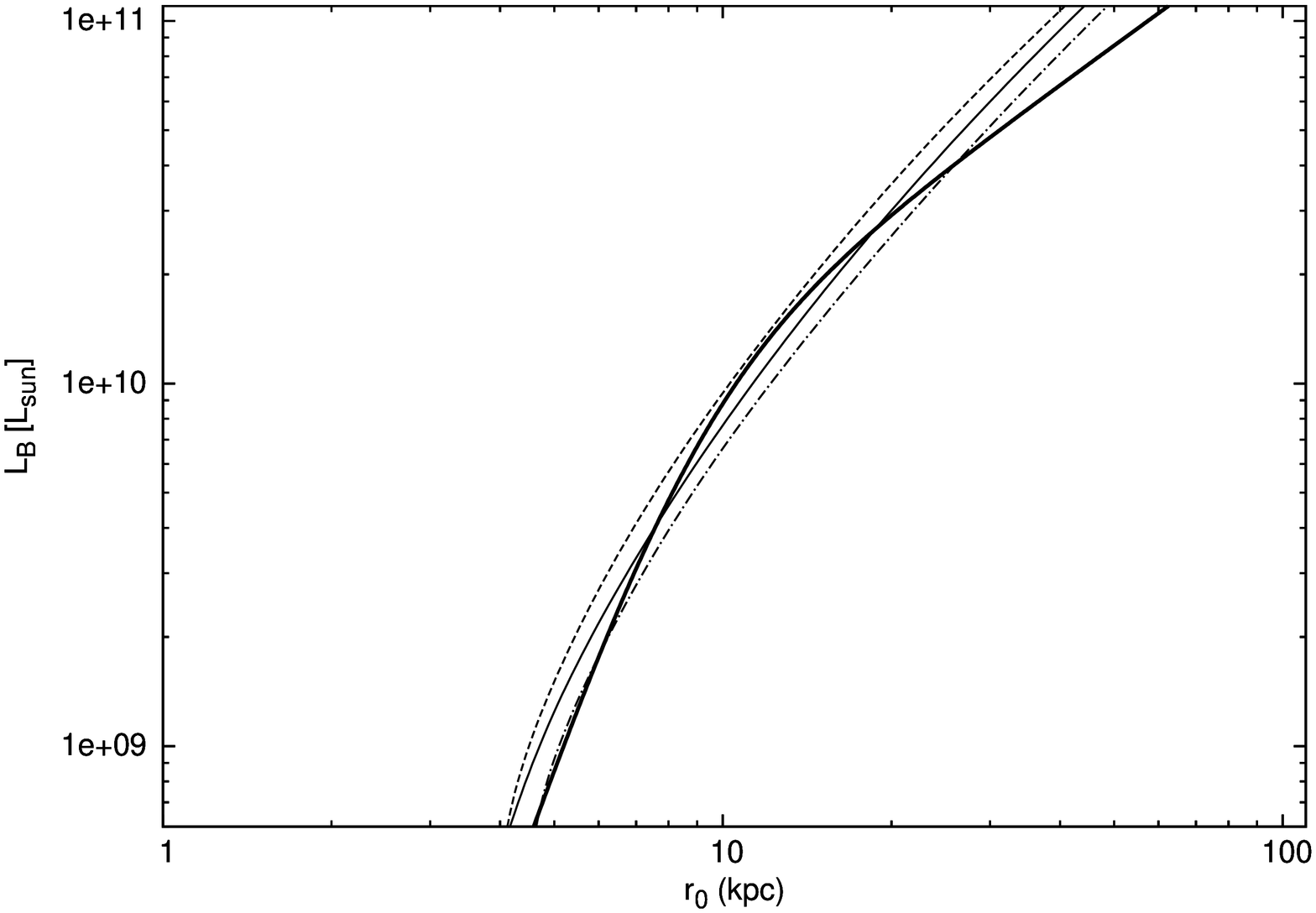, angle=270, width=12.5cm}}
\vskip 0.2cm
\noindent
{\small Figure 7a: 
Derived galaxy luminosity versus core radius $r_0$
for various values of $c_1$ which parameterize the hardness
of the supernova mirror photon spectrum. 
Plotted are: 
$c_1 = 1$ (dashed line), 
$c_1 = 2$ (solid line) and $c_1 = 3$ (dashed-dotted line).
Also shown (thick solid line) is the corresponding empirical galactic scaling relation obtained
from Eq.(\ref{rel1}).
}
\vskip 0.1cm
\centerline{\epsfig{file=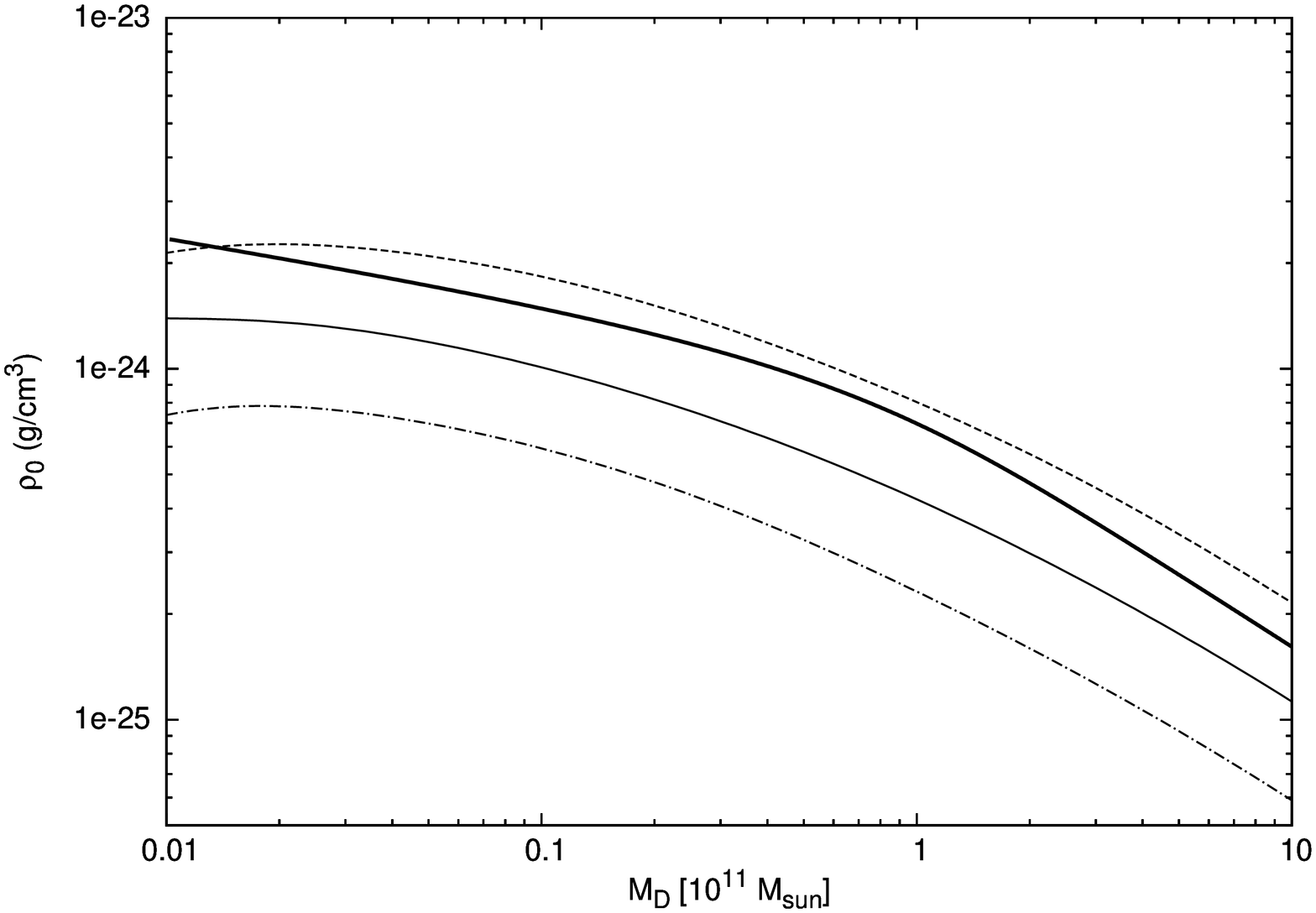, angle=270, width=12.4cm}}
\vskip 0.2cm
\noindent
{\small Figure 7b: 
Derived dark matter central density,
$\rho_0$, versus baryonic mass, $m_D$
for various values of $c_1$. 
Plotted are: 
$c_1 = 1$ (dashed line), 
$c_1 = 2$ (solid line) and $c_1 = 3$ (dashed-dotted line).
Also shown (thick solid line) is the corresponding empirical galactic scaling relation obtained
from Eq.(\ref{mdrel}) and Eq.(\ref{rel1}).
}
\vskip 0.1cm

\centerline{\epsfig{file=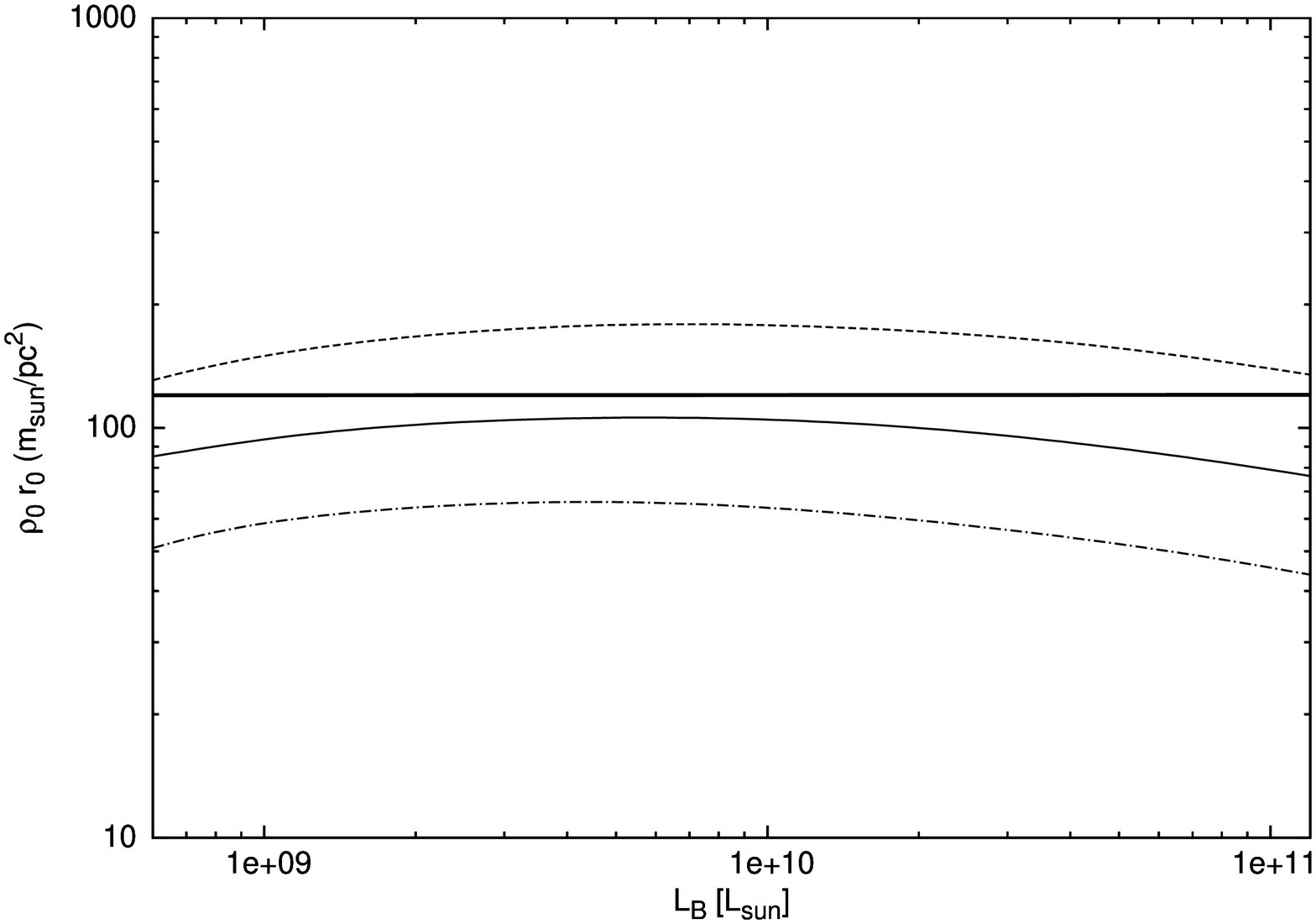, angle=270, width=12.4cm}}
\vskip 0.2cm
\noindent
{\small Figure 7c: 
Derived $\rho_0 r_0$ as a function of galactic luminosity, $L_B$
for various values of $c_1$.
Plotted are: 
$c_1 = 1$ (dashed line), 
$c_1 = 2$ (solid line) and $c_1 = 3$ (dashed-dotted line).
Also shown is $\rho_0 r_0 = 120 \ m_{\odot}/{\rm pc}^2$,
close to the central value of the `empirical' relation
Eq.(\ref{rel1}) (thick solid line).
}
\vskip 1.0cm

The figures demonstrate that the derived relations for $\rho_0, r_0$ are compatible with the
rough `empirical' scaling relations given in Eq.(\ref{rel1})\footnote{We have neglected any
possible scaling difference between the galactic $r^*$ band luminosity, $L_r$, and B-band 
luminosity, $L_B$.}. 
The level of agreement seems to be very nontrivial. The only parameter adjusted was 
${L'}_{SN}^{MW}$ which was set so that $\rho_0 r_0$ had a  value $\sim 10^2 m_\odot/{\rm pc}^2$
for $m_D = 10^{11} m_\odot$,  $c_1 = 2$.

To check the robustness of these results
with respect 
to reasonable parameter 
variations we have varied $\xi_{Fe'}, \ {L'}_{SN}^{MW}$. 
[Note that changing $E_c$ has the same effect as changing ${L'}_{SN}^{MW}$ and is
therefore not considered.]
In figure 8 [figure 9] 
we show the effect of varying ${L'}^{MW}_{SN}$ [$\xi_{Fe'}$], 
with the other parameters unchanged. 
These figures 
demonstrate that
the variation of each of these parameters by an order of magnitude around our reference
values does {\it not} greatly modify the $L$ versus $r_0$ relation (figure 7a).
They also show that
$\rho_0 r_0$  is still constant but the value of the constant is modified somewhat.
We have also found that the above conclusions hold also when 
$f \equiv n_{H'}/n_{He'}$, is changed.
Thus, the scaling properties demonstrated in figures 7 remain valid even when parameters
are varied. 
Figures 7,8,9 indicate that the effect of 
varying $c_1, \ \xi_{Fe'}$ and ${L'}_{SN}^{MW}$ on $\rho_0 r_0$ can be roughly approximated by:
\begin{eqnarray}
\rho_0 r_0 \simeq \left[ {\xi_{Fe'} \over 0.02}\right]^{0.7} \ 
\left[ { {L'}_{SN}^{MW} \over 2\times 10^{45} \ {\rm erg/s}} \right]^{0.7}
\left[ {2 \over c_1}\right] 
\ 10^{2} \ m_\odot/{\rm pc}^2
\ .
\label{40}
\end{eqnarray}
On the other hand the $L$ versus $r_0$ relation is remarkably insensitive 
to variations of the parameters. 
Instead of fixing the baryonic parameters via the relations, Eq.(\ref{rd}) and Eq.(\ref{mdrel}),
we can consider independent variations of $m_D, \ L$ and $r_D$. In doing so, we find that
$r_0$ is mainly set by the parameter $r_D$.  That is, it is primarily the disk scale length
that sets the scale for the dark matter core radius.  
This suggests that the primary relation for $r_0$ is one in terms of $r_D$ (not $L$).
Fixing the baryonic parameters as per Eq.(\ref{rd}) and Eq.(\ref{mdrel}), and minimizing $\Delta$
we obtain the approximate numerical result:
\vskip -0.5cm
\begin{eqnarray}
r_0 \approx 3.0 \ \left( {r_D \over {\rm kpc}}\right)^{1.1} \ {\rm kpc} 
\ .
\end{eqnarray}
As with our other results, this is valid over the 
considered baryonic mass range of spirals 
$10^{9} m_\odot \stackrel{<}{\sim} m_D \stackrel{<}{\sim} 10^{12} m_\odot$, and 
assumed the Burkert profile. 



These results are all very interesting, and among other things, support the premise that
kinetic mixing is likely close to $\ \epsilon = 10^{-9}$.
Such a value has already been identified as a region of interest from
the analysis\cite{foot} of direct detection experiments, such as DAMA\cite{dama}. 
The latter can be explained with $\epsilon \sqrt{\xi_{Fe'}} \approx 2\times 10^{-10}$.
Observe that since ${L'}_{SN}^{MW} \propto \epsilon^2$ [Eq.(\ref{raf1x})], 
Eq.(\ref{40}) suggests
\begin{eqnarray}
\rho_0 r_0 \sim \left[ {\epsilon \sqrt{\xi_{Fe'}} \over 5\times 10^{-10}}\right]^{1.4} \ 10^2 \ m_\odot/{\rm pc}^{2}
\ .
\end{eqnarray}
This demonstrates the compatibility of the galactic scaling relations with results from the
direct detection experiments.

\centerline{\epsfig{file=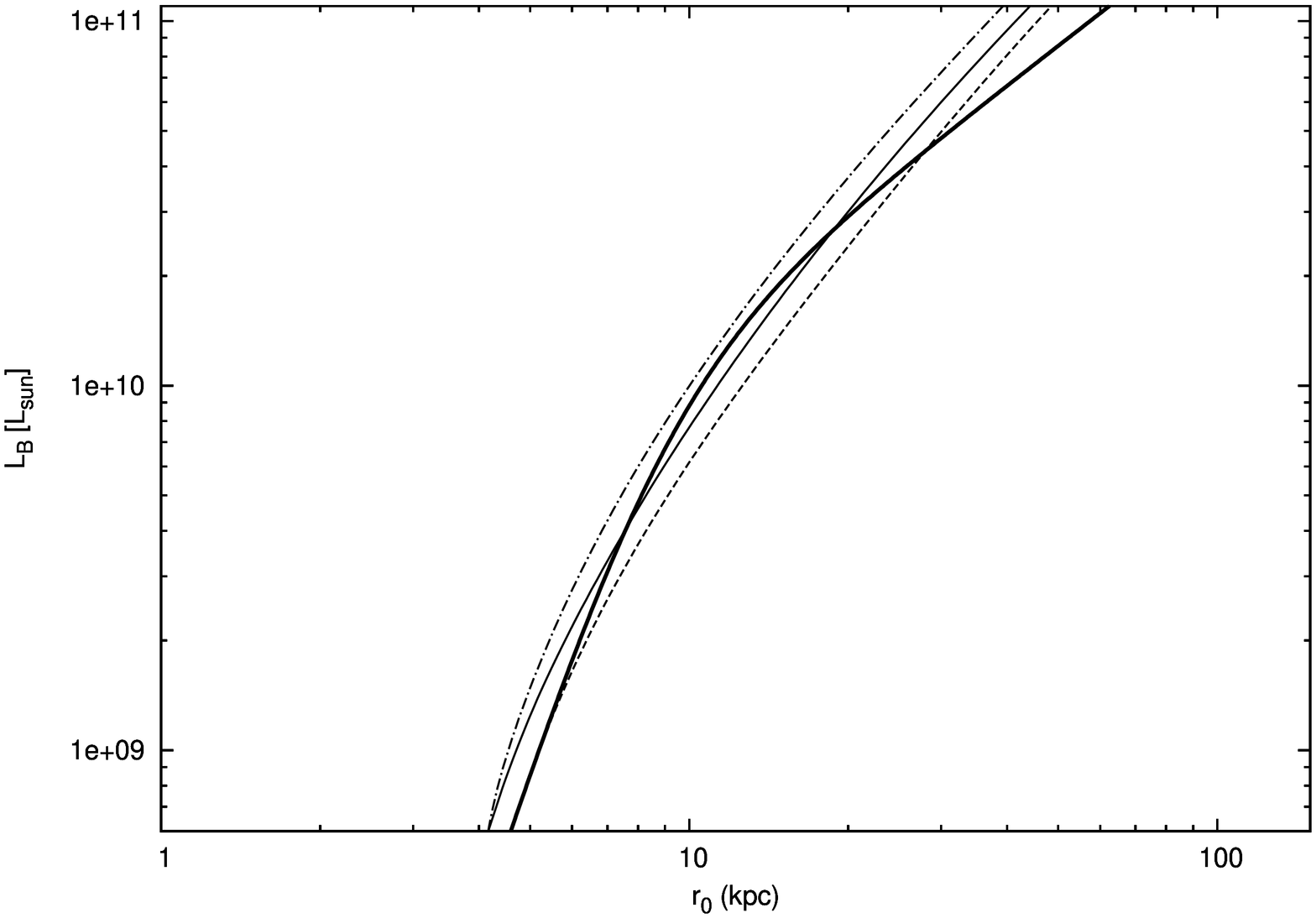, angle=270, width=11.8cm}}
\vskip 0.2cm
\noindent
{\small Figure 8a: 
Derived galaxy luminosity versus core radius $r_0$ 
for $c_1 = 2$ and various values of ${L'}_{SN}^{MW}$. 
Plotted are ${L'}_{SN}^{MW} = 0.7 \times 10^{45}$ erg/s (dashed line) 
${L'}_{SN}^{MW} = 2.0 \times 10^{45}$ erg/s (solid line) 
and 
${L'}_{SN}^{MW} = 6.0 \times 10^{45}$ erg/s (dashed-dotted line). 
Also shown (thick solid line) is the corresponding empirical galactic scaling relation obtained
from Eq.(\ref{rel1}).
}
\vskip 0.1cm

\centerline{\epsfig{file=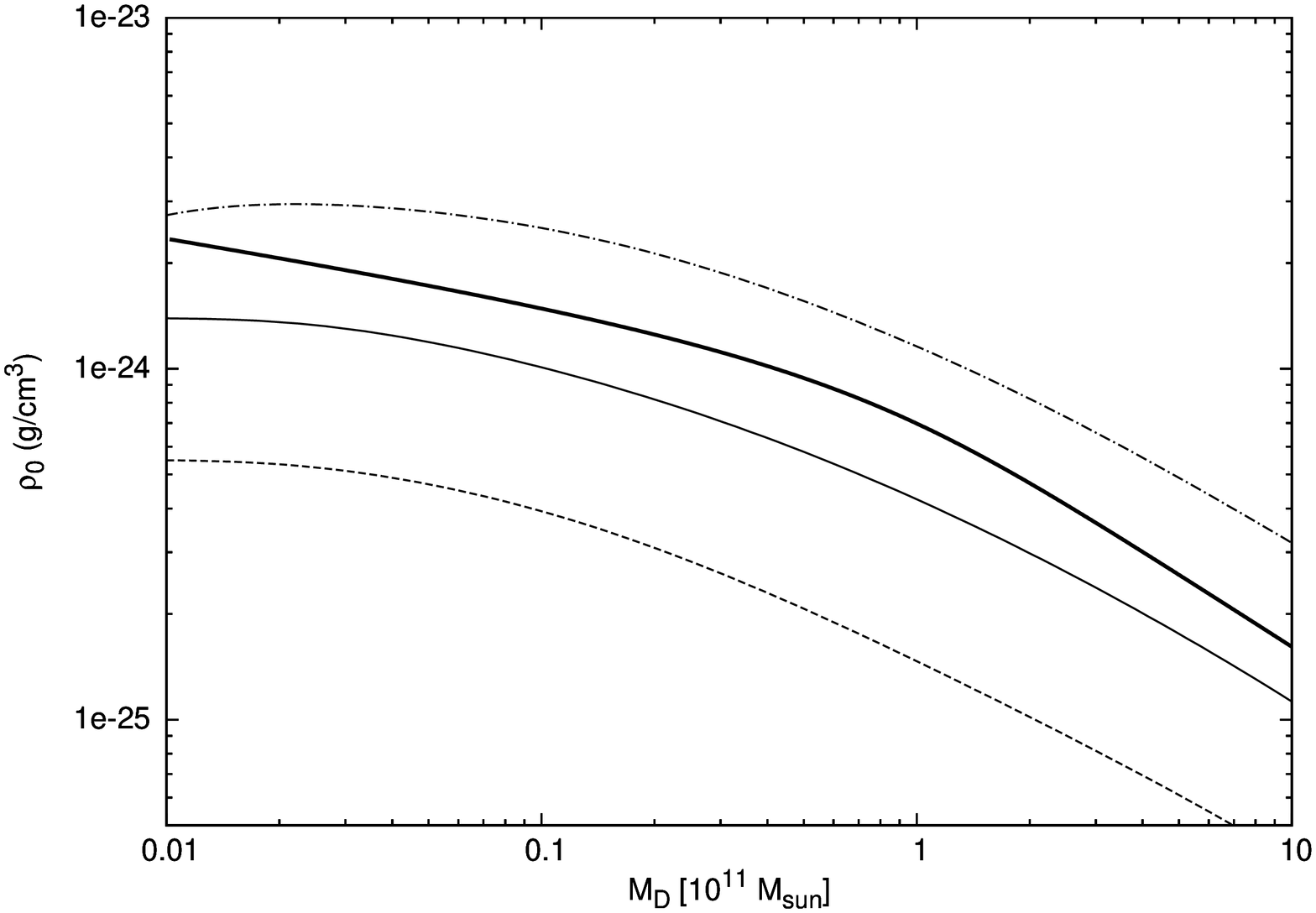, angle=270, width=12.0cm}}
\vskip 0.2cm
\noindent
{\small Figure 8b: 
Derived dark matter density,
$\rho_0$, versus baryonic mass, $m_D$
for $c_1 = 2$ and various values of ${L'}_{SN}^{MW}$: 
${L'}_{SN}^{MW} = 0.7 \times 10^{45}$ erg/s (dashed line) 
${L'}_{SN}^{MW} = 2.0 \times 10^{45}$ erg/s (solid line) 
and 
${L'}_{SN}^{MW} = 6.0 \times 10^{45}$ erg/s (dashed-dotted line). 
Also shown (thick solid line) is the corresponding empirical galactic scaling relation obtained
from Eqs.(\ref{mdrel},\ref{rel1}).
}
\vskip 0.1cm
\centerline{\epsfig{file=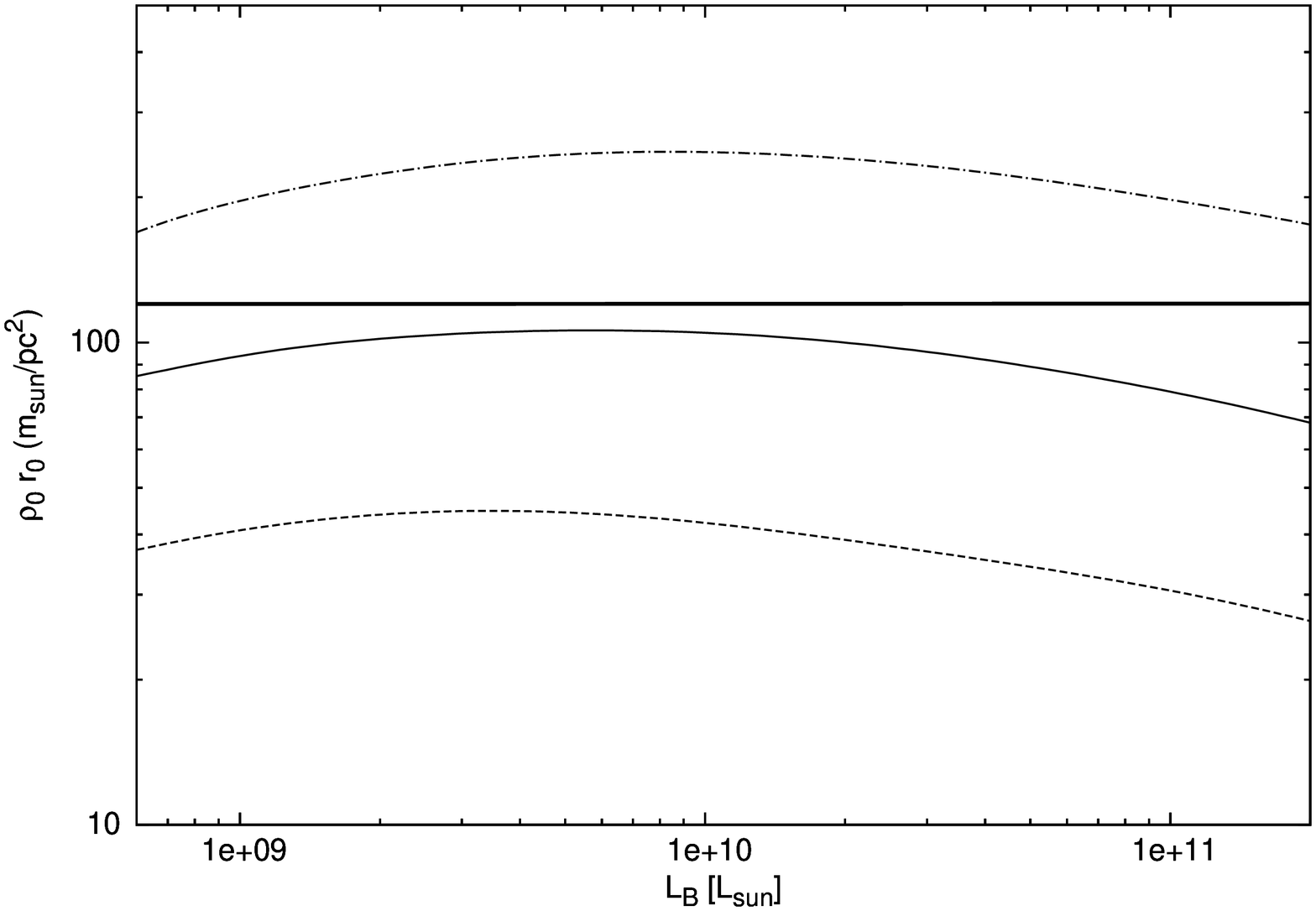, angle=270, width=12.0cm}}
\vskip 0.3cm
\noindent
{\small Figure 8c: 
Derived
$\rho_0 r_0$ as a function of galactic luminosity, $L_B$
for $c_1 = 2$ and various values of ${L'}_{SN}^{MW}$. 
Plotted are ${L'}_{SN}^{MW} = 0.7 \times 10^{45}$ erg/s (dashed line) 
${L'}_{SN}^{MW} = 2.0 \times 10^{45}$ erg/s (solid line) 
and 
${L'}_{SN}^{MW} = 6.0 \times 10^{45}$ erg/s (dashed-dotted line). 
Also shown is $\rho_0 r_0 = 120 \ m_{\odot}/{\rm pc}^2$,
close to the central value of the `empirical' relation
Eq.(\ref{rel1}) (thick solid line).
}

\centerline{\epsfig{file=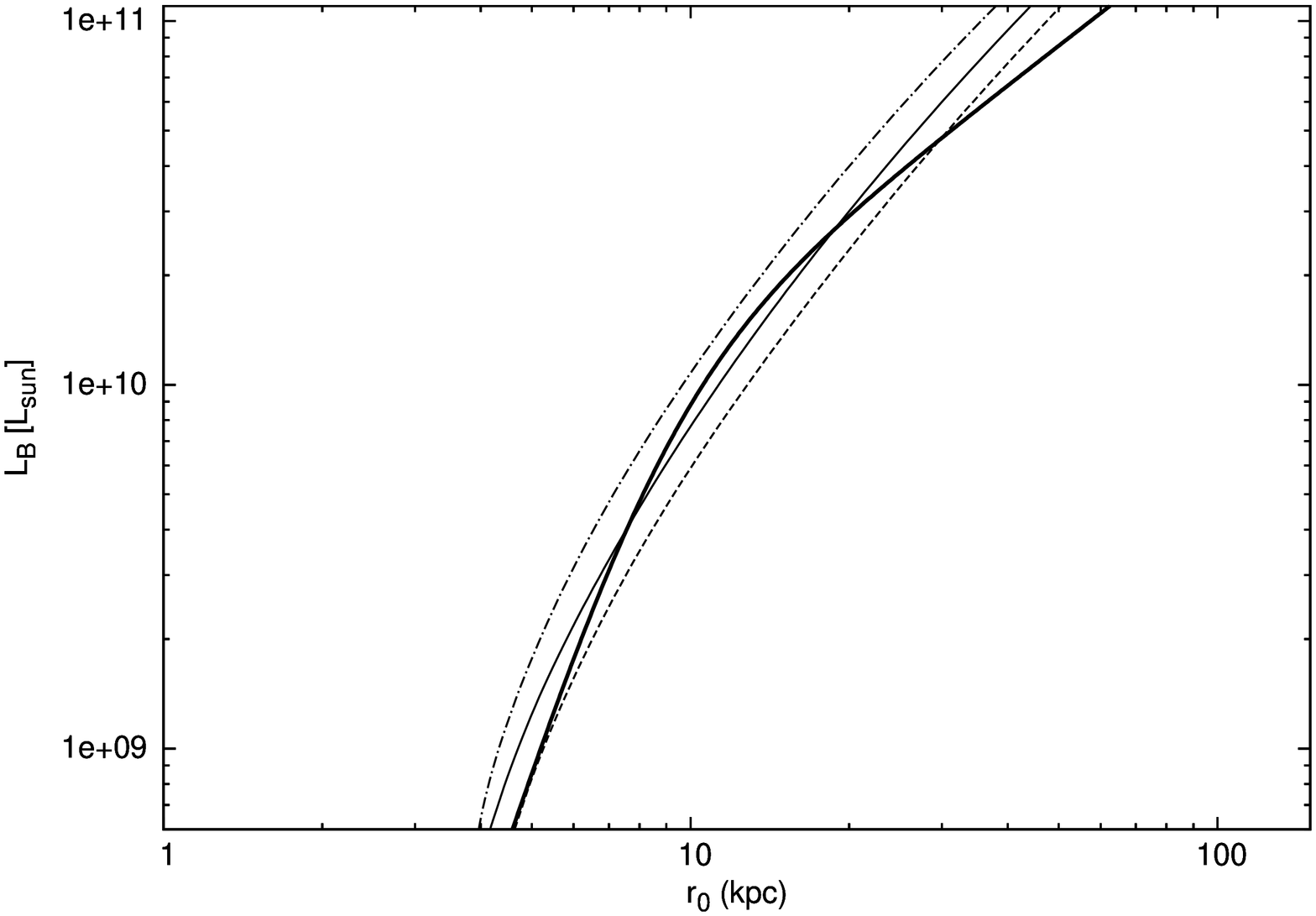, angle=270, width=12.3cm}}
\vskip 0.2cm
\noindent
{\small Figure 9a: 
Derived galaxy luminosity versus core radius $r_0$ 
for $c_1 = 2$ and various values of 
$\xi_{Fe'}$. 
Plotted are $\xi_{Fe'} = 0.007$ (dashed line), 
$\xi_{Fe'} = 0.02$ (solid line) 
and
$\xi_{Fe'} = 0.06$ (dashed-dotted line). 
Also shown (thick solid line) is the corresponding empirical galactic scaling relation obtained
from Eq.(\ref{rel1}).
}
\vskip 0.1cm

\centerline{\epsfig{file=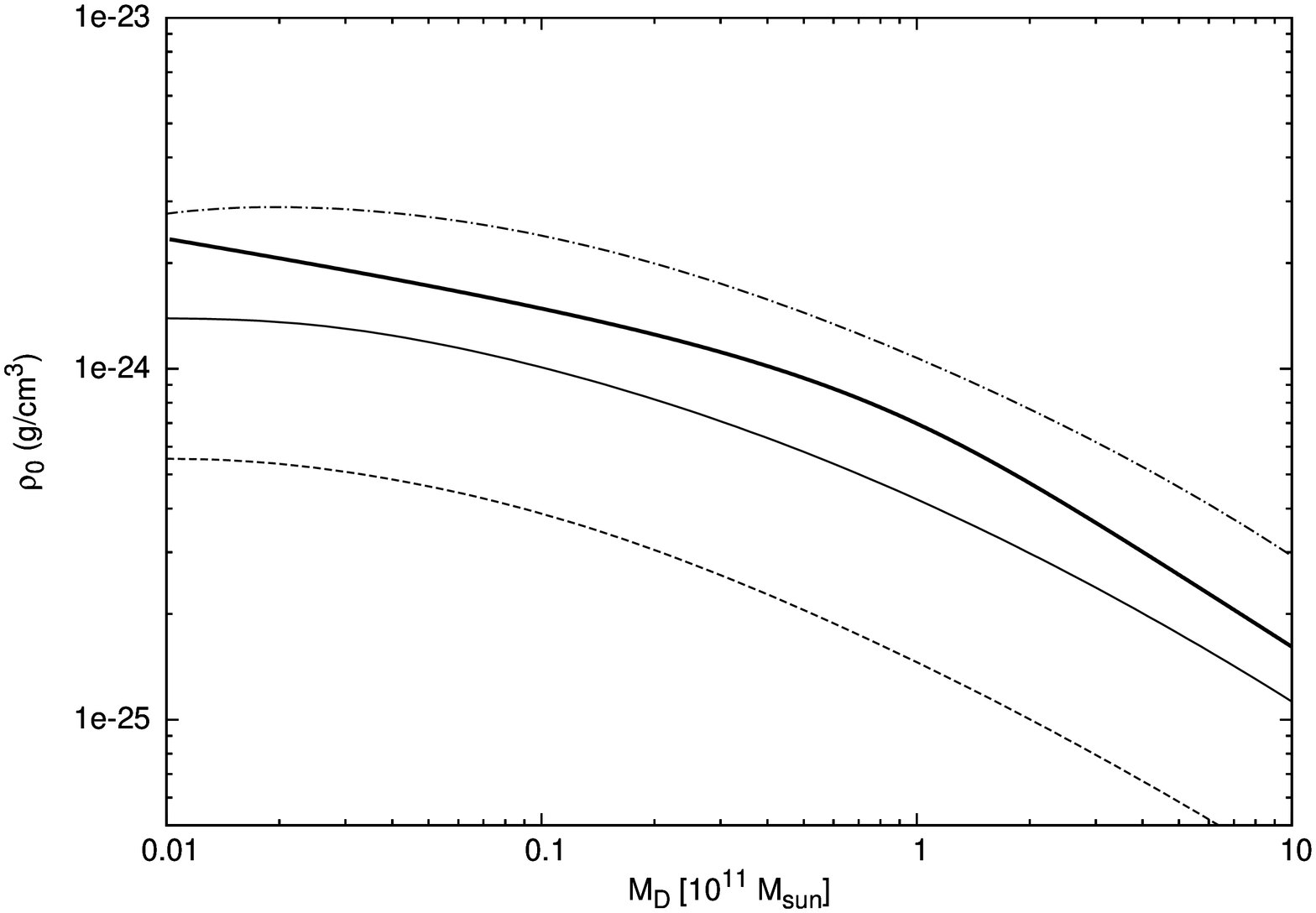, angle=270, width=12.0cm}}
\vskip 0.2cm
\noindent
{\small Figure 9b: 
Derived dark matter central density,
$\rho_0$, versus baryonic mass, $m_D$
for $c_1 = 2$ and various values of 
$\xi_{Fe'}$. 
Plotted are $\xi_{Fe'} = 0.007$ (dashed line), 
$\xi_{Fe'} = 0.02$ (solid line) 
and
$\xi_{Fe'} = 0.06$ (dashed-dotted line). 
Also shown (thick solid line) is the corresponding empirical galactic scaling relation obtained
from Eqs.(\ref{mdrel},\ref{rel1}).
}
\vskip 0.1cm
\centerline{\epsfig{file=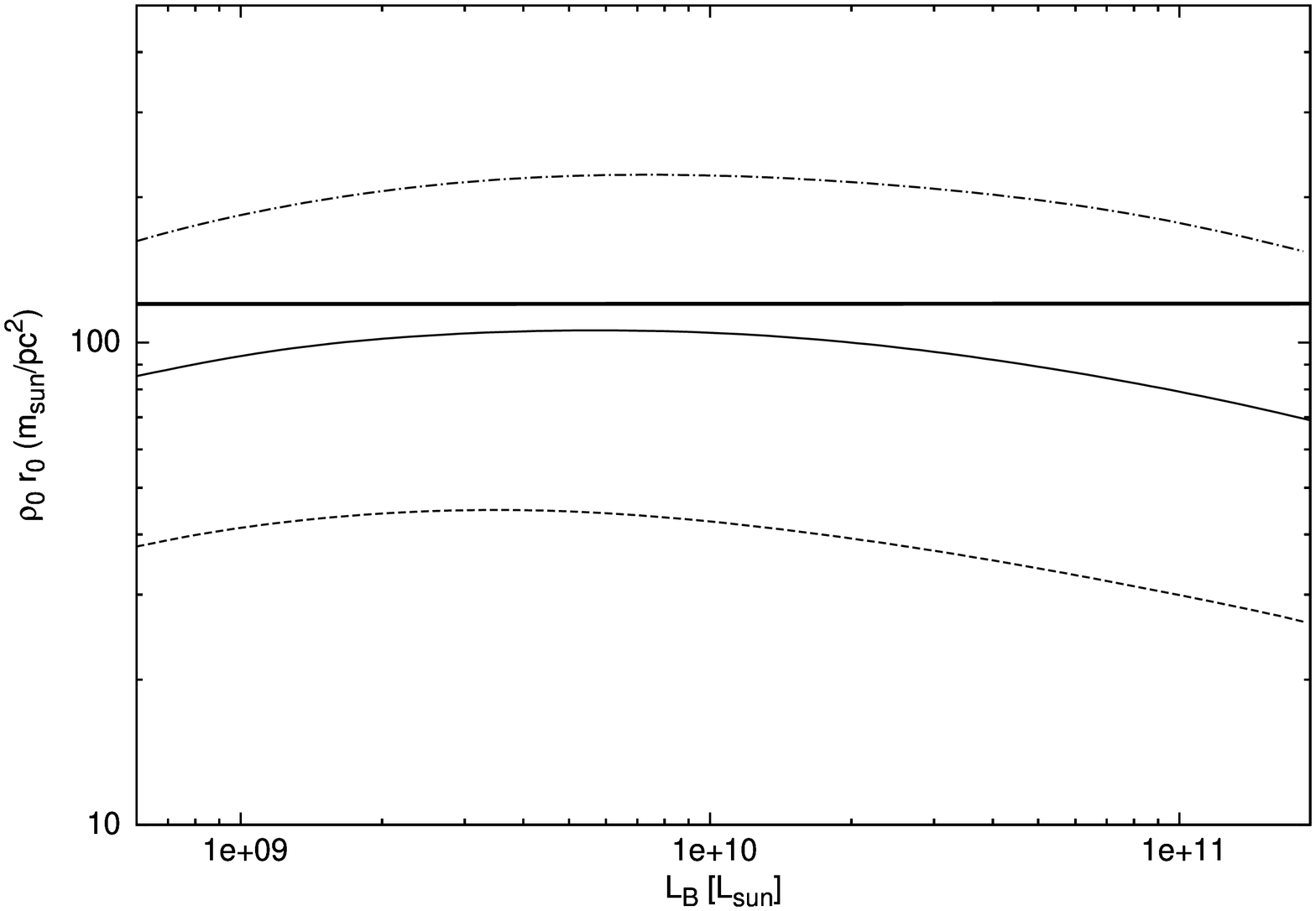, angle=270, width=12.0cm}}
\vskip 0.3cm
\noindent
{\small Figure 9c: 
Derived
$\rho_0 r_0$ as a function of galactic luminosity, $L_B$
for $c_1 = 2$ and various values of 
$\xi_{Fe'}$. 
Plotted are $\xi_{Fe'} = 0.007$ (dashed line), 
$\xi_{Fe'} = 0.02$ (solid line) 
and
$\xi_{Fe'} = 0.06$ (dashed-dotted line). 
Also shown is $\rho_0 r_0 = 120 \ m_{\odot}/{\rm pc}^2$,
close to the central value of the `empirical' relation
Eq.(\ref{rel1}) (thick solid line).
}


\section{Dwarf spheroidal galaxies}

The results of the previous section are very encouraging. One 
might be tempted to investigate other classes
of galaxies, i.e. galaxies beyond spirals, and inquire if they are 
also compatible within this dissipative dark matter picture. 
An interesting class of galaxies is the  dwarf spheroidal galaxies.
These are much smaller than spirals and feature
luminosity around $\sim 10^5-10^6 L_\odot$. Observations indicate that these
galaxies are much more dark matter dominated 
than spirals.

The cooling rate in dwarf spheroidal galaxies is suppressed if the halo mirror plasma 
temperature is below around  3 eV.
For such temperatures figure 2 indicates that
most of the mirror helium would be in neutral atoms.
Only the smaller mirror hydrogen subcomponent is ionized 
and could thereby participate in bremsstrahlung cooling.  

\centerline{\epsfig{file=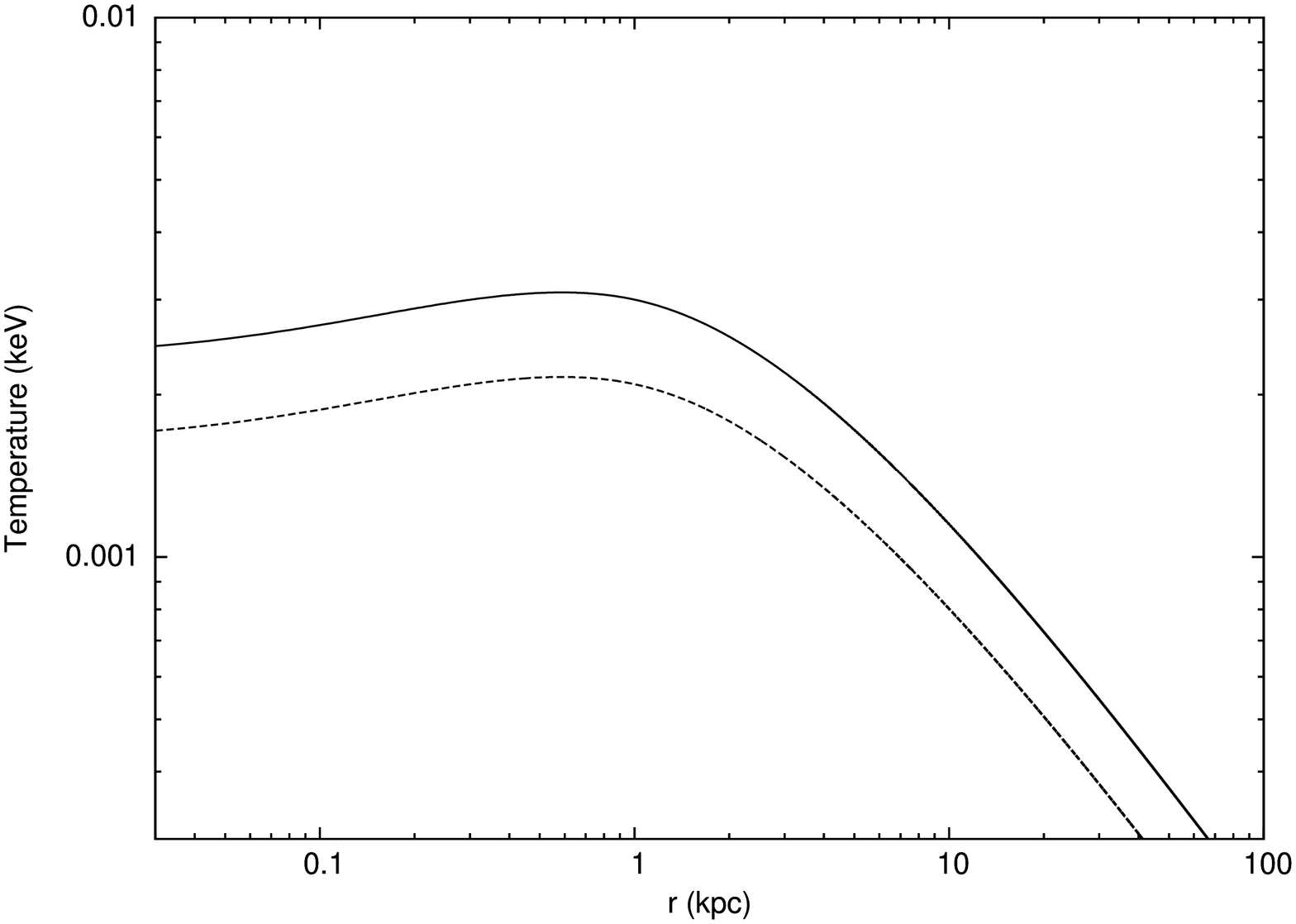, angle=270, width=12.8cm}}
\vskip 0.4cm
\noindent
{\small Figure 10: The mirror plasma temperature profile  for a dwarf spheroidal galaxy.
Shown is an example with
dark matter core radius $r_0 = 0.5$ kpc  
and central density $\rho_0 = 10^{-23} \ {\rm g/cm}^3$ (solid line). 
Also shown is another example with
the same dark matter core radius $r_0 = 0.5$ kpc  
but slightly higher central density $\rho_0 = 5\times 10^{-24} \ {\rm g/cm}^3$ (dashed line). 
}

\vskip 1.2cm

The halo temperature for dwarf spheroidal galaxies can be computed in the
same way in which we computed the temperature for spirals, discussed in section 3.
Consider two examples with
(a) dark matter core radius $r_0 = 0.5$ kpc and 
central density $\rho_0 = 10^{-23}\ {\rm g/cm^3}$ and  
(b) $r_0 = 0.5$ kpc and
central density $\rho_0 = 5\times 10^{-24} \ {\rm g/cm}^3$. 
Both examples are
consistent with the 
observations\cite{saluccixxx}.
The results of numerically solving Eq.(\ref{p9}) are shown in figure 10.
When computing the temperature profile for
dwarf spheroidal galaxies, we have neglected
the baryon component completely as far as its contribution to the local acceleration, $g$.
This is reasonable because the baryonic mass is estimated to be only a few percent 
of the total mass for these galaxies\cite{saluccixxx}.
Note that for both examples, figure 2 indicates that only a small
proportion of mirror helium will be fully ionized.
[Numerically we find that example (a) has 
mean mass parameter given by $\bar m \simeq 1.6$ GeV, while example (b)
has $\bar m \simeq 2.3$ GeV.]


Importantly dwarf spheroidal galaxies appear to have relatively little ordinary gas
component and do not exhibit current star formation\cite{grebel}.
Thus, for dwarf spheroidal galaxies it does not seem possible to stabilize 
a spherical mirror plasma component with energy from ordinary supernova.
Some other energy source would be needed or possibly, the mirror dark matter has
collapsed into a disk/bulge component for this galaxy class.

\section{Elliptical galaxies and clusters}

Elliptical galaxies are another interesting class of objects to think about.
As with dwarf spheroidal galaxies,
elliptical galaxies appear to be largely
devoid of gas and do not
exhibit significant star formation at the present epoch. 
Thus the current rate 
of ordinary supernovae is very low.
In the absence of supernova heating,
the mirror particle plasma is 
expected to undergo gravitational collapse
onto a disk/bulge. For large elliptical galaxies, 
the time scale for this to occur can be long, and even today,
many such galaxies might not be fully collapsed. 
In any case, in the absence of a significant 
heat source it is reasonable to expect 
the dark matter within elliptical galaxies to be flattened out, to some extent,
due to dissipative processes.
This picture is 
consistent, at least qualitatively, with the observed ellipticity 
of elliptical galaxies, e.g.\cite{chand}.
That is, for mirror dark matter, and closely related hidden sector models, the non-spherical distribution of the
dark matter in elliptical galaxies can potentially be explained  
due to  dissipation effects (in the absence of which, 
the observed ellipticity could have been used to place stringent limits on
dark matter self-interactions, as in e.g.\cite{feng}).

Within larger structures, such as clusters of galaxies, mirror dark matter
self-interactions might lead to important effects.
In particular, observations of the bullet cluster
have been used to argue for stringent constraints on dark matter self-interactions\cite{clowe}.
Recall, the bullet cluster is an example of a system in which a 
collision between two clusters has apparently taken place.
Each cluster has three components, the galaxies, hot intergalactic gas and then there is the dark matter.
When the main cluster and subcluster collide, the hot (ordinary matter) gas associated with the two colliding clusters appears to be
slowed, but not stopped, by interactions. Both the galaxies and dark matter components appear to pass through each other.  
These observations pose a potential puzzle for mirror dark matter. Why doesn't the dark matter within each cluster slow
down due to interactions?
Part of the explanation could be that the fraction 
of dark matter existing as hot gas unbound to individual galaxies (i.e. intergalactic gas) might be less than the corresponding fraction  
for ordinary matter\cite{sil99}.
Another part of the explanation could be that mirror dark matter has
self-interactions which are weaker than those of ordinary matter due to environmental conditions. In particular, if the mirror 
particle plasma has higher temperature than the ordinary particle plasma.
Recall that
the mirror electron scattering cross-section behaves like $d\sigma/d\Omega \propto 1/v^4 \sim 1/T^2$. 
The temperature, $T$, of the mirror particle plasma is uncertain, but the hydrostatic equilibrium condition, Eq.(\ref{p9}),
suggests that $T \propto \bar m$.  
The higher helium mass fraction in the mirror sector means that $\bar m \approx $ 1.1 GeV for mirror dark matter, 
cf. $\bar m \approx 0.55$ GeV for ordinary matter, and thus the temperature of the mirror plasma 
in clusters might be higher than that of the ordinary plasma by a factor of around two or so.
If this is indeed part of the explanation of the bullet cluster observations, 
then studies of colliding clusters which feature lower temperatures
could find an offset between the galaxy component and total mass distribution. 
Interestingly, there are some tentative hints in this 
direction\cite{hints}.

\section{Concluding remarks}

We have examined galactic structure within the context of dissipative dark matter 
candidates, focusing on the mirror dark matter model.
At first sight, dissipative dark matter might seem unlikely,
given the inferred approximate sphericity of dark matter halo's in spiral galaxies.
However, the idea\cite{sph} that there exists a heat source to counteract the halo cooling
due to dissipative interactions and that this heat source could be ordinary
supernova seems to be possible.
At any rate,
it is a specific idea, and mirror dark matter offers a specific model in which to study it.
 
In this article we have provided a detailed numerical analysis
of this whole picture.
Although a number of assumptions are made, and a few corners may have been cut, our results are very encouraging.
Our analysis indicates that the inferred dark matter scaling properties of spiral galaxies, discovered
by Salucci and others, are explicable within this dark matter framework.
Moreover our results are remarkably insensitive to many of the unknown parameters of the theory
(such as details of the supernova energy spectrum, mirror metal mass fraction, etc).
Thus, we have reason to be confident that this dissipative dark matter  picture,
and to some extent the specific mirror dark matter implementation of it,  may well be on the
right track in explaining galactic structure.
That is, we agree with the title of ref.\cite{core}, ``Dark matter in galaxies: leads to its nature''.


\vskip 1.9cm
\noindent
{\large \bf Acknowledgements}

\vskip 0.2cm
\noindent
This work was supported by the Australian Research Council.

\end{document}